\documentclass[fleqn,10pt]{wlscirep}
\usepackage[utf8]{inputenc}
\usepackage[T1]{fontenc}
\usepackage{dcolumn}
\usepackage{textcomp}

\usepackage{multirow}
\usepackage{dcolumn}
\usepackage{textcomp}
\usepackage{longtable}
\usepackage{graphicx}
\usepackage{caption}
\usepackage{booktabs}

\usepackage{xcolor}

\title{Perceiving exposure segregation with open urban imagery}

\author[1,*]{Yunke Zhang}
\author[1,*]{Ruolong Ma}
\author[1]{Xin Zhang}
\author[1]{Yu Shang}
\author[1]{Fengli Xu}
\author[2]{Tong Xia}
\author[1,**]{Yong Li}
\affil[1]{Department of Electronic Engineering, Beijing National Research Center for Information Science and Technology (BNRist), Tsinghua University, Beijing, China}
\affil[2]{Vanke School of Public Health, Tsinghua University, Beijing, China}
\affil[*]{These authors contributed equally.}
\affil[**]{corresponding authors: Yong Li (liyong07@tsinghua.edu.cn)}

\begin{abstract}
Socioeconomic exposure segregation -- the lack of daily interaction between income groups -- erodes social capital and entrenches inequality, yet the specific physical features that drive these behavioral restrictions remain poorly understood. Prior research has quantified where segregation occurs using mobility data, but has not identified how the built environment facilitates or inhibits these interactions. Here we introduce VISAGE, a large multi-modal model-enabled framework that perceives exposure segregation directly from open satellite and street-level imagery across 10,030 communities in 31 U.S. cities. Moving beyond black-box correlations, we operationalize cross-disciplinary sociological theory into an interpretable visual codebook to detect physical regulators of social mixing. We find that the built environment encodes a legible grammar of segregation: "defensible" architectural forms (e.g., fences, gated enclosures) and monofunctional zoning systematically predict higher social isolation, whereas mixed-use infrastructure fosters interaction, explaining substantial variance in mobility-derived segregation patterns (Pearson $r=0.770$). Crucially, we show that inclusionary housing policies manifest in distinct visual signatures associated with higher mixing, suggesting that policy interventions successfully alter the physical landscape to encourage diversity. Our findings offer a scalable pathway to decipher the social production of space, providing a mechanism-based lens to understand how the built environment shapes social behavior.
\end{abstract}
\begin{document}

\flushbottom
\maketitle
\thispagestyle{empty}

\section*{Main}

Cities, now home to over half the global population, are often envisioned as engines of interaction, yet access to urban spaces and opportunities remains highly uneven~\cite{florida2017new}. One visible and consequential manifestation is socioeconomic segregation, the spatial and social separation of income groups that limits inter-class contact~\cite{chetty2014land, chetty2022social}, entrenches mobility barriers~\cite{yang2025urban}, and distorts perceptions of inequality~\cite{davidai2024economic}. Crucially, recent work has shifted attention to \textit{exposure segregation}~\cite{moro2021mobility, athey2021estimating, nilforoshan2023human, abbiasov202415, liao2025effect} -- a behavioural metric defined by the extent to which daily activity spaces are shared across income groups. This shift moves beyond static, residence-based measures (\emph{e.g.}, dissimilarity and isolation~\cite{duncan1955methodological, massey1988dimensions}) to capture dynamics shaped by human mobility and the built environment~\cite{xu2025using, arvidsson2023urban, liao2025socio, yabe2025behaviour, han2025visitation}. Yet, investigating the physical determinants of exposure segregation at scale remains elusive. While mobility data quantifies where segregation occurs, it is frequently access-restricted and disconnected from the fine-grained environmental context of why it occurs~\cite{gallotti2024distorted, yabe2024enhancing}. To bridge this gap, open satellite and street-level imagery is increasingly used to characterize neighborhood attributes~\cite{fan2023urban, gebru2017using, naik2014streetscore}. However, current visual approaches remain largely correlation-based and anchored to narrow, study-specific assumptions. By contrast, exposure segregation is driven by interacting mechanisms across housing, transportation, land use, and social space -- mechanisms that demand interpretable, theory-driven measurement rather than purely predictive proxies~\cite{naik2017computer, crawford2015aesthetic, juhasz2023amenity, francis2012creating, qi2024understanding}. Systematically validating these theoretical mechanisms across vast urban landscapes requires a methodological shift beyond simple visual pattern matching.

This gap between theoretical complexity and observational capacity calls for a methodological paradigm shift toward knowledge-integrated computational frameworks capable of operationalizing abstract social concepts at scale. In parallel, Large Language Models (LLMs) have evolved to synthesize vast, fragmented domain literature -- from urban planning to environmental psychology -- transforming qualitative insights into structured, testable hypotheses~\cite{delgado2025transforming, zhang2025exploring}. Simultaneously, Large Multi-Modal Models (LMMs) have advanced beyond rigid image classification to semantic reasoning, enabling the interpretation of complex urban scenes through language-grounded explanations rather than opaque feature vectors~\cite{hou2025urban, lin2024vila, schulze2025visual}. Leveraging these developments offers a unique opportunity to decipher the physical laws of exposure segregation: (i) by distilling cross-disciplinary theories into a compact, interpretable ``visual codebook'' of built environment determinants; (ii) by using the semantic reasoning capabilities of LMMs to detect these subtle cues in open imagery, effectively scaling expert-level observation to entire metropolitan regions~\cite{zhang2025urbanmllm, zhang2025perceiving}; and (iii) by operating a closed-loop workflow in which held-out empirical evidence iteratively refines these theoretical definitions, ensuring that measurement is driven by validated physical mechanisms rather than spurious correlations.

We operationalize this paradigm in VISAGE, the first large multi-modal model-enabled framework designed formechanism-driven urban sensing using only open imagery (Figure~\ref{fig:framework}). VISAGE is structured as a knowledge-integrated workflow that automates the loop of sociophysical discovery regarding exposure segregation. It begins by synthesizing domain literature to distill a concise, theory-informed ``visual codebook'' of physical cues --from fence heights to zoning types -- explicitly linked to mechanisms of socioeconomic mixing. This codified knowledge is then converted into structured reasoning templates, serving as mechanism-aware supervision signals. Finally, a domain-adapted LMM detects these cues in imagery and reasons from scene semantics to infer exposure segregation. Crucially, this approach utilizes out-of-sample evaluation feedback to iteratively refine its codebook hypotheses, ensuring that the resulting measure is transparent, auditable, and grounded in sociological knowledge rather than simple visual correlation, providing a robust blueprint for studying how the physical environment shapes human social behavior at scale.

VISAGE contributes three fundamental advances to the study of urban inequality. First, it establishes scalable generalizability by demonstrating that open satellite and street-view imagery can reliably proxy exposure segregation across 10,030 communities in 31 major U.S. metropolitan areas (Pearson $r=$0.770). This enables consistent, high-resolution measurement of social mixing even in regions where mobility data is unavailable. Second, it ensures methodological transparency by anchoring predictions to a theory-grounded codebook, moving the field from opaque ``black-box'' correlations to mechanism-informed auditing. Finally, and most significantly, our framework unravels novel physical mechanisms of socioeconomic isolation. We empirically establish that high segregation is systematically enforced by ``defensive'' urban forms -- such as high residential fences and monofunctional zones -- while interaction is fostered by specific configurations like mixed-use diversity and accessible public spaces. These verified patterns identify actionable physical levers for urban planning, offering a rigorous method to evaluate policy interventions such as inclusionary housing~\cite{wang2023inclusionary}. Together, these results illustrate how knowledge-integrated AI can reframe urban sensing from simple observation to automated sociophysical discovery, marking a substantive leap forward in the computational understanding of social behavior~\cite{hou2025urban}.
\section*{Results}

\subsection*{Deriving a theoretical visual grammar of exposure segregation}

\begin{figure*}[htbp]
    \centering
    \includegraphics[width=\linewidth]{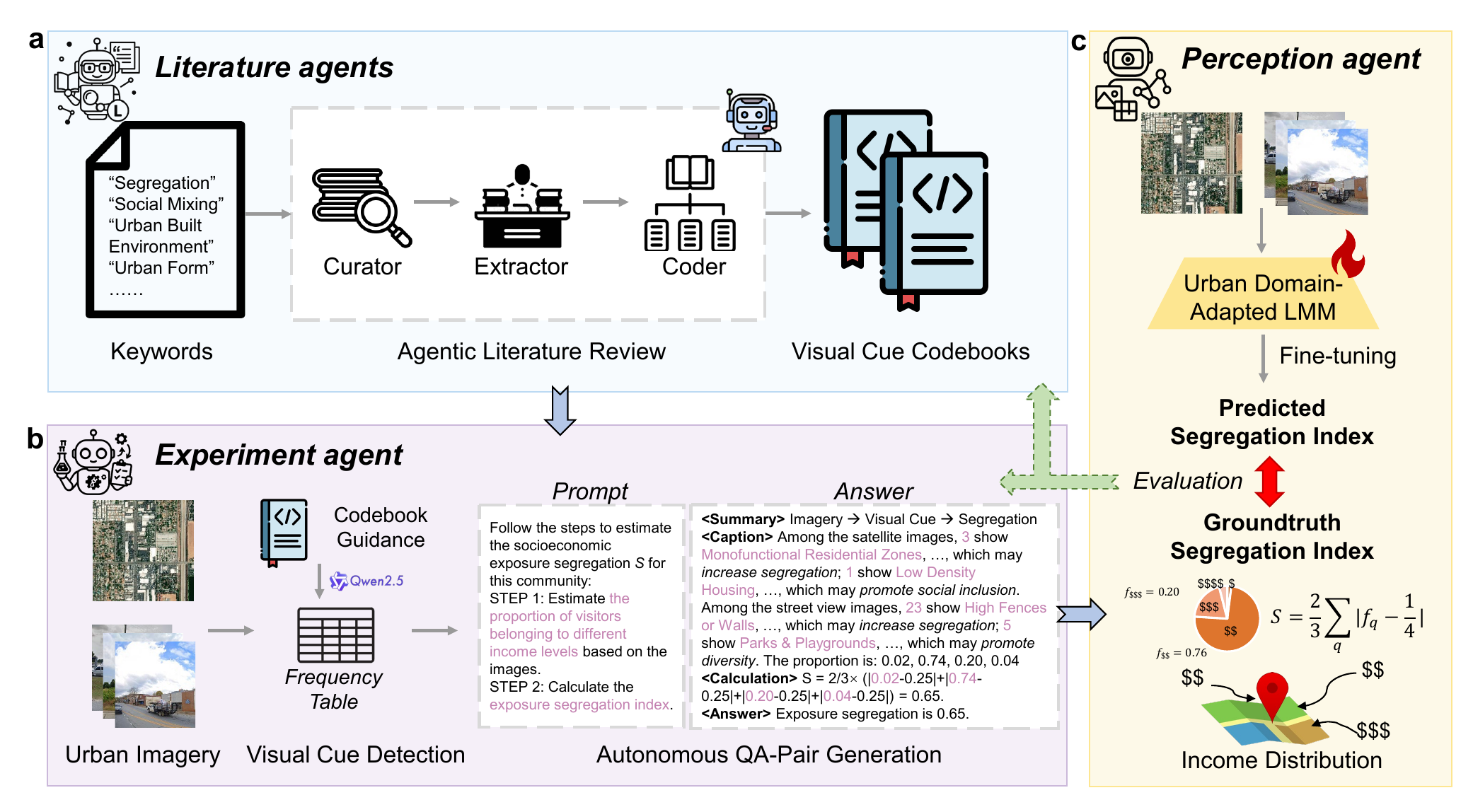}
    \caption{\textbf{VISAGE: The knowledge-integrated framework for perceiving exposure segregation.} \textbf{a}, LLMs acting as Literature Agents curate peer-reviewed sources to distill a theory-informed visual codebook, identifying image-observable attributes explicitly linked to mechanisms of socioeconomic mixing or separation. \textbf{b}, The Experiment Agent translates these theoretical cues into observable variables by detecting them in paired satellite and street-level imagery. It aggregates these detections into community-level evidence and generates codebook-aware reasoning scaffolds (summary, caption, calculation, answer) to guide the model. \textbf{c}, A domain-adapted LMM acting as the Perception Agent fuses these visual cues to infer visitor income composition and the exposure-segregation index. An iterative feedback loop uses out-of-sample mobility data to update the codebook hypotheses, ensuring the final visual grammar is empirically validated against human behavioral patterns. See Methods M2 and Supplementary Notes S1 for details.}
    \label{fig:framework}
\end{figure*}

To systematically investigate how the built environment shapes socioeconomic mixing, we developed VISAGE, a knowledge-integrated framework organized as a three-stage agentic workflow (Figure~\ref{fig:framework}). Unlike traditional ``black-box" deep learning that relies on opaque features, this system operationalizes abstract sociological theories into quantifiable visual evidence through a closed-loop discovery process.

Exposure segregation is not driven by single pixels, but by complex configurations of urban form. To capture this, the \textit{Literature Agent} first synthesizes a broad spectrum of peer-reviewed literature -- spanning urban sociology, transportation planning, and environmental psychology -- into a compact, interpretable visual codebook (Figure~\ref{fig:framework}a). Acting as a knowledge distiller, this agent screens diverse studies to identify specific physical features hypothesized to influence social mixing. For instance, it identifies that ``defensible space'' theories link physical barriers (e.g., High Fences, Gated Entrances) to reduced interaction~\cite{LeGoix2005GatedCommunities, Vesselinov2008MembersOnly}, while ``New Urbanist'' theories posit that Mixed-Use Frontages foster diverse encounters~\cite{Fan2023Diversity, WangVermeulen2021LifeBetweenBuildings}. This process yields a structured ontology of 44 image-observable cues (Supplementary Tables S1), explicitly linking visual attributes to the mechanism of socioeconomic segregation. Methods M2 and Supplementary Figure S1-3 provide workflow details.

To test these theoretical mechanisms at scale, the \textit{Experiment agent} converts the visual codebook into structured reasoning protocols that guide the interpretation of urban imagery. Rather than simply classifying images, this agent uses a multi-modal detector~\cite{bai2025qwen2} to identify specific codebook-defined cues in paired satellite and street-view imagery (Figure~\ref{fig:framework}b). It then auto-generates a chain-of-thought (CoT) rationale following a four-part structure~\cite{xu2024llava}. For every community, it constructs a question-answer pair that asks the model to first identify physical evidence (e.g., ``presence of security bars'') and then synthesize these observations to infer visitor income composition. This structured approach anchors predictions in observable physical evidence, preventing reliance on spurious correlations and ensuring that the inferred segregation is mechanistically linked to the built environment.

Finally, the \textit{Perception Agent} -- a domain-adapted LMM designed to fuse complementary signals from cross-view imagery~\cite{zhang2025urbanmllm} -- performs the critical task of validating these hypotheses against ground truth (Figure~\ref{fig:framework}c). By processing the structured cues and imagery, it infers the exposure segregation index and provides an auditable explanation. Crucially, the workflow employs an iterative validation loop where the \textit{Perception Agent} utilizes out-of-sample feedback from mobility data to prune visual cues that fail to consistently predict segregation. This data-driven refinement ensures that the final codebook (Supplementary Tables S2-3) represents a verified ``visual grammar'' of segregation, retaining only those physical elements—from infrastructure barriers to amenity diversity—that demonstrate a robust, empirical link to human social behavior.

\subsection*{Open imagery proxies exposure segregation}

\begin{figure*}[htbp]
    \centering
    \includegraphics[width=\linewidth]{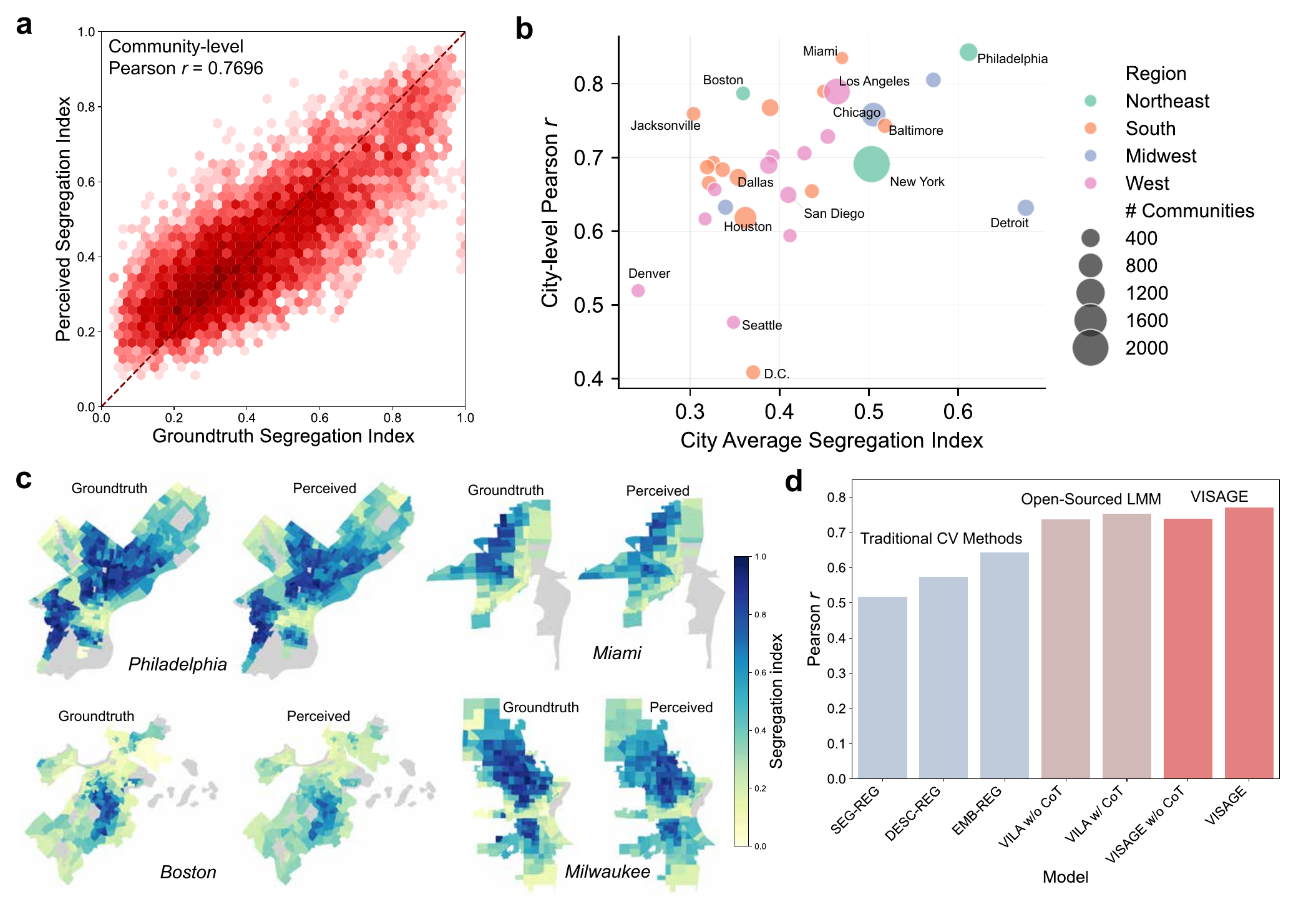}
    \caption{
        \textbf{Reliability of VISAGE in perceiving exposure segregation.} \textbf{a}, Community-level scatter of perceived versus groundtruth segregation index for 10,030 communities across 31 U.S. cities. The red dashed line represents perfect prediction. \textbf{b}, Relationship between the city-level Pearson correlation coefficient ($r$) and the city's mean segregation index. Dot size corresponds to the number of communities in each city. Colors indicate the city's geographical region within the country. \textbf{c}, Spatial distributions of the groundtruth and VISAGE-perceived segregation index for communities in Philadelphia, Miami, Boston, and Milwaukee. \textbf{d}, Comparison of Pearson $r$ on held-out test sets across various baseline approaches. These include (i) conventional computer vision (CV) baselines utilizing semantic segmentation, caption-derived features, or generic embeddings, and (ii) open-source LMM baselines, alongside (iii) the VISAGE framework. All methods were evaluated using identical experimental setups to ensure a fair comparison.
    }
    \label{fig:evaluation}
\end{figure*}

We evaluated the reliability of VISAGE in proxying socioeconomic exposure segregation across the 31 largest U.S. metropolitan areas. Exposure segregation is defined as an index $S \in [0,1]$ that summarizes the diversity of daily encounters across four city-specific income quartiles observed in a community~\cite{moro2021mobility} (Figure~\ref{fig:framework}c). $S=0$ denotes perfectly balanced mixing across quartiles, while $S=1$ indicates complete homogeneity. Groundtruth $S$ is computed for 10,030 communities using aggregated, de-identified device visits from 2019 (SafeGraph Weekly Patterns), a dataset widely established for quantifying cross-group exposure~\cite{nilforoshan2023human, zhang2024counterfactual}. For each community, we inferred the income distribution of visitors and computed $S$ using Equation~\ref{equ:seg} (Methods M1). The study footprint spans diverse urban morphologies across the Northeast, Midwest, South, and West, ranging from historic cores with fine-grained blocks to post-war suburban expansions (Supplementary Figure S4). Reliability is assessed on held-out, disjoint splits using the Pearson correlation $r$ between predicted and groundtruth $S$.

At the community scale, the framework's image-based estimates track mobility-derived groundtruth closely (Figure~\ref{fig:evaluation}a). The association is strong and approximately linear over the full range of $S$ (Pearson $r$=0.7696 across 10,030 communities), indicating that multi-view scene semantics extracted from paired street-view and satellite images contain sufficient signal to recover fine-grained variation in exposure segregation. This result validates the efficacy of the knowledge-integrated workflow: rather than relying on black-box visual features, open imagery is first summarized into theory-informed evidence via the codebook, allowing the domain-adapted LMM to reason from observable mechanisms to a community-level estimate.

Aggregating to the city scale shows that reliability remains high across heterogeneous contexts (Figure \ref{fig:evaluation}b). Several metropolitan areas exceed $r$>0.80, including Philadelphia ($r$=0.8428), with Miami and Milwaukee also above this threshold. A positive cross-city trend links higher average segregation to higher reliability ($r$=0.4360, $p$=0.142), suggesting that where segregation gradients are more pronounced, the corresponding visual signals—such as physical barriers or sharp zoning transitions—are more distinct. Conversely, in cities with more uniform social mixing, between-community variation is smaller, making prediction a finer-grained discrimination task. These patterns indicate that VISAGE generalizes across distinct regional morphologies while reflecting real differences in the strength of visual evidence for $S$.

Within cities, VISAGE reproduces both macro-structures and local contrasts (Figure~\ref{fig:evaluation}c and Supplementary Figure S9). In Philadelphia, Miami, Boston, and Milwaukee, the perceived surfaces recover broad clusters and gradients visible in groundtruth and preserve fine-scale differences across adjacent communities. Areas with more uniform values appear as continuous regions, while sharp transitions align with recognizable urban boundaries. The close correspondence between the two surfaces argues that the model learns organized spatial patterns rooted in the built environment rather than isolated artifacts.

We further contextualize this performance by comparing VISAGE with representative alternatives operating on the same imagery. These include conventional computer-vision pipelines based on semantic-segmentation features, caption-derived features, and generic embeddings, as well as open-source LMMs adapted to this task (Figure~\ref{fig:evaluation}d and Supplementary Figure S8; see Methods M2 and Supplementary Note S2.4 for details). Under identical preprocessing, aggregation, and splits, VISAGE attains the highest reliability.

Across 31 metropolitan areas and 10,030 communities, VISAGE yields reliable perception of exposure segregation from open satellite and street-level imagery, reproduces the salient spatial organization observed in groundtruth, and maintains performance across diverse urban forms. These results demonstrate that a theory-informed framework, which combines sociological evidence with multi-view reasoning, achieves deeper, auditable urban perception than correlation-based prediction alone.

\subsection*{Physical mechanisms of socioeconomic isolation}

\begin{figure*}[htbp]
    \centering
    \includegraphics[width=\linewidth]{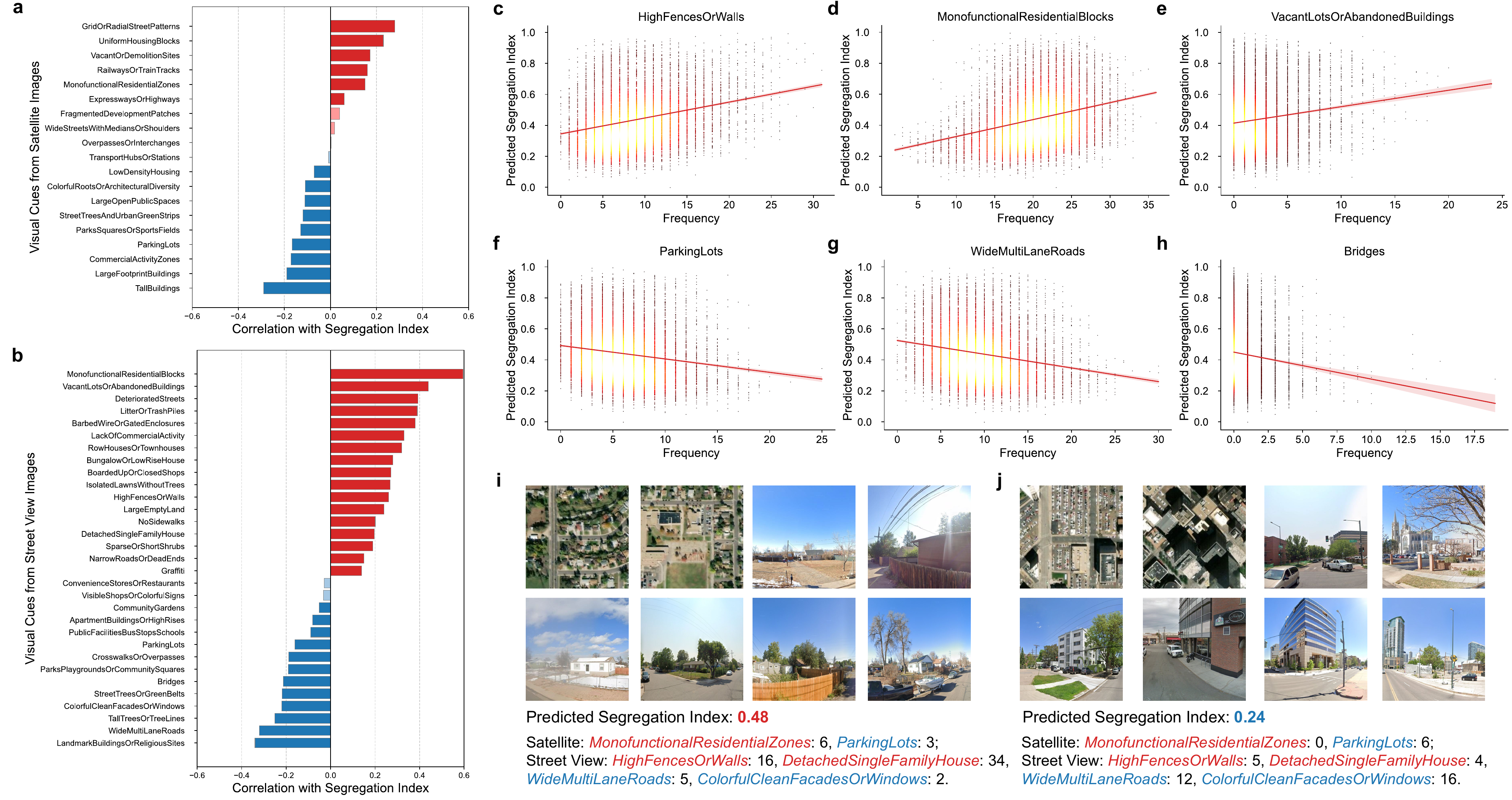}
    \caption{
        \textbf{VISAGE reasons exposure segregation from visual cues.} \textbf{a}, Correlations between satellite image visual cue frequencies and groundtruth segregation across 10,030 communities. \textbf{b}, Correlations between street-view image visual cue frequencies and groundtruth segregation. \textbf{c-h}, Response curves illustrating how representative cues vary with the predicted segregation index; examples include positively associated cues (\textbf{c-e}) and negatively associated cues (\textbf{f-h}). Lines show mean trends; shaded bands indicate 95\% CIs. \textbf{i-j}, Contrasting case studies. Left: a community with higher predicted segregation (example cue mix includes Detached Single-Family Houses and Monofunctional Residential Zones; few cues associated with mixing). Right: a community with lower predicted segregation (prevalence of Wide Multi-Lane Roads and Colorful, Clean Facades; few high-barrier cues). Full cue set and correlations are provided in Supplementary Figure S10-S11.}
    \label{fig:explain}
\end{figure*}

Beyond establishing predictive reliability, our framework enables the discovery of the specific physical mechanisms that drive exposure segregation. By auditing the system's visual codebook, we investigate whether the theoretical features distilled from literature—such as defensible space or land-use diversity—empirically correlate with mobility-based isolation, and how these cues are composed to explain segregation patterns.

We first validate the theoretical codebook against ground truth by correlating the frequency of every visual cue with the exposure-segregation index across all communities (Figure \ref{fig:explain}a-b; excluding cues with $p$>0.05). The analysis reveals a distinct physical dichotomy. The strongest positive associations with segregation are observed for exclusionary features: \textit{High Fences or Walls}, \textit{Barbed Wire or Gated Enclosures}, and \textit{Monofunctional Residential Blocks}. Conversely, the strongest negative associations (promoting mixing) are linked to accessible infrastructure: \textit{Large Open Public Spaces}, \textit{Street Trees}, \textit{Commercial Activity Zones}, and \textit{Wide Multi-Lane Roads}. These empirical directions align perfectly with the codebook's theoretical expectations, confirming that "boundary" and "vacancy" signals systematically accompany higher segregation, while "openness" and "activity" facilitate interaction. Critically, this result validates the efficacy of the cross-disciplinary synthesis: it demonstrates that visual hypotheses spanning sociology, urban planning, and transportation 1 can be effectively integrated to identify the structural impediments to social mixing at a scale exceeding typical human-curated hypothesis testing.

We then examined how these mechanisms are synthesized in the model’s reasoning process. For each community, the chain-of-thought rationale links the cue table to a prediction through a fixed structure: identifying key cues, interpreting them via codebook directions, and composing them into a segregation estimate $S$. Response curves derived from the model’s outputs vary smoothly with cue prevalence and follow the expected signs (Figure \ref{fig:explain}c-h). For instance, the prevalence of \textit{Vacant Lots or Abandoned Buildings} shows a positive relation with predicted segregation ($r$=0.1562), whereas \textit{Wide Multi-Lane Roads} shows a negative relation ($r$=-0.2171). This alignment between empirical correlations, response curves, and the explanation text indicates that predictions arise from composing multiple mechanistic cues rather than from single-feature shortcuts.

Two representative cases illustrate how these cue configurations drive the physical production of segregation (Figure \ref{fig:explain}i-j). A low-segregation community is characterized by a combination of \textit{Wide Multi-Lane Roads} and \textit{Colorful Clean Facades} with minimal enclosure elements. The reasoning highlights these as indicators of permeability and commercial activity that foster diverse encounters. mainly. In contrast, a high-segregation community is defined by the co-occurrence of \textit{Detached Single-Family Houses} and \textit{Monofunctional Residential Blocks} alongside boundary cues, which the model attributes to a ``defensible'' form that restricts inter-group contact.

Taken together, these analyses provide an interpretable, mechanism-based account of exposure segregation. The results consolidate a broad sociophysical picture: enclosure and monofunctional residential forms are physically linked to higher social isolation, whereas openness, greenery, and mixed-use diversity are linked to lower segregation. By recovering these patterns from open imagery, the framework not only matches prior theoretical expectations but clarifies mechanism-level regularities at a national scale, advancing our understanding of how the built environment physically regulates exposure segregation.

\subsection*{Impacts of social mixing policies on exposure segregation}

\begin{figure*}[htbp]
    \centering
    \includegraphics[width=\linewidth]{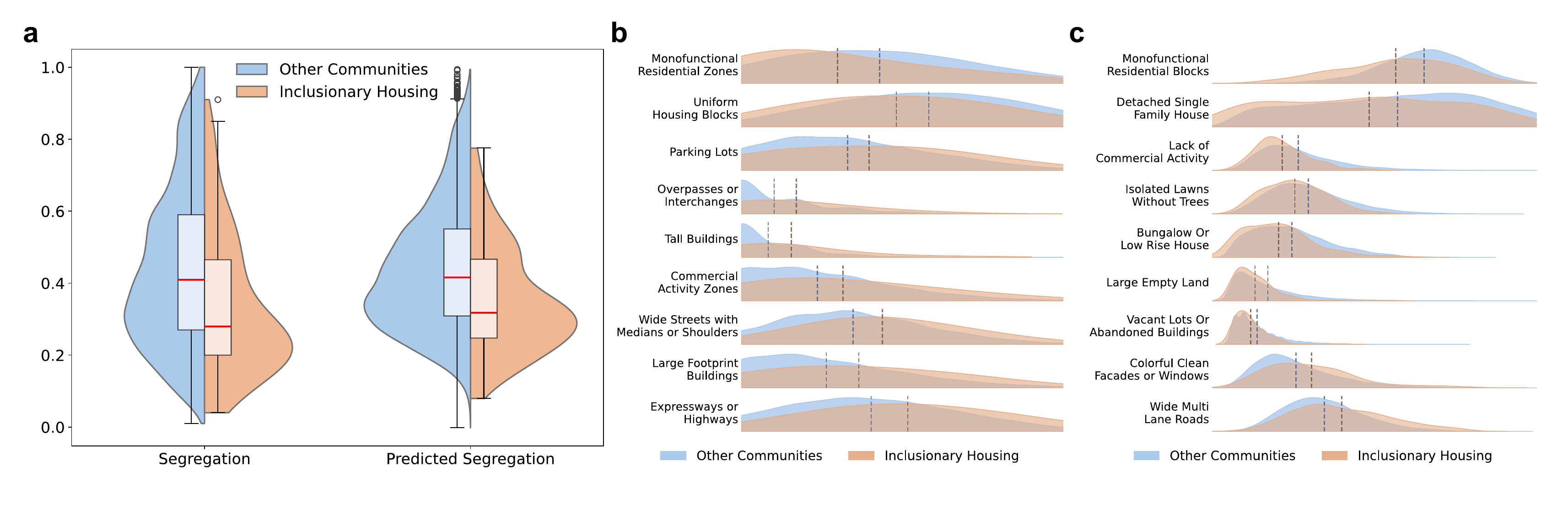}
    \caption{
        \textbf{Visual signatures of inclusionary housing policies.} \textbf{a}, Group comparison of exposure segregation between communities with and without inclusionary housing programs; boxes show median and interquartile range (IQR); whiskers are 1.5IQR; two-sided t-tests are conducted on groundtruth $S$ and predicted $\hat{S}$. \textbf{b-c}, Differences in satellite (\textbf{b}) and street-view (\textbf{c}) cue prevalence between inclusionary housing and non-inclusionary housing communities (top cues by absolute difference; points are mean differences with 95\% CIs).
    }
    \label{fig:inclusive}
\end{figure*}

Socioeconomic exposure segregation is not simply a spatial statistic. It is a lived reality shaped by the policies, investments, and design choices that structure urban life. Among the interventions designed to address segregation, inclusionary housing (IH) policies -- requiring or incentivizing the integration of affordable units into new residential developments -- are often cited as promising tools for fostering social mixing~\cite{azlan2025inclusionary}. While earlier studies have assessed their effects on housing supply and affordability, much less is known about whether these policies actually alter the exposure segregation of urban residents, and if so, how such changes are manifested in the physical environment~\cite{zhang2024housing, thurber2018spatially}.

To systematically assess whether these interventions leave a measurable footprint on the urban fabric, we analyzed 171 communities identified as having implemented inclusionary housing according to the Inclusionary Housing Map \& Program Database~\cite{wang2023inclusionary}. Comparing mobility-derived outcomes reveals a significant disparity: communities with active IH policies exhibit a groundtruth segregation index of 0.3345, markedly lower than the 0.4433 observed in non-policy communities (two-sample t-test, $p < 0.001$; Figure~\ref{fig:inclusive}a). Crucially, this pattern is mirrored in our framework's image-based predictions (0.3529 vs 0.4413, $p < 0.001$). Regression analyses controlling for racial composition, median income, poverty rate, and other demographic covariates confirm robustness: inclusionary housing communities maintain a segregation index lower by $-0.0684 \pm 0.015$ ($p < 0.001$) and a predicted index lower by $-0.0482 \pm 0.011$ relative to comparable non-policy neighborhoods (Supplementary Table S4). These results indicate that social mixing policies are indeed associated with lower levels of both measured and perceived segregation, consistent with their intended role in fostering interaction across income groups.

We further audited the specific visual cues to understand how these policies reshape the built environment. As shown in Figure~\ref{fig:inclusive}b-c, inclusionary housing communities exhibit a distinct visual signature: they show a significantly higher prevalence of \textit{Wide Multi-Lane Roads}, \textit{Colorful Clean Facades}, and \textit{Commercial Activity Zones}, alongside a lower prevalence of \textit{Monofunctional Residential Zones} and \textit{Empty Land}. These patterns suggest that IH policies do not merely change resident demographics but actively promote mixed-use building forms, active public spaces, and a diversity of commercial functions—physical configurations that theoretically catalyze socioeconomic integration. The fact that our framework captures these distinctions indicates that the detected reductions in segregation are rooted in substantive, observable changes to the physical production of space, rather than superficial artifacts.

Taken together, these findings validate the framework's ability to recover theory-consistent associations between visual cues and exposure segregation at scale. By linking policy interventions to specific shifts in the built environment—from monofunctional zoning to mixed-use activity—the analysis demonstrates that automated urban sensing can serve as a rigorous instrument for auditing urban governance. The workflow is readily extensible to other design interventions, offering a scalable pathway to monitor how planning decisions physically shape the social fabric of cities.
\section*{Discussion}

VISAGE marks a substantive paradigm shift in urban sensing by reformulating socioeconomic exposure segregation from a latent mobility statistic into a physically observable attribute of urban form. Historically, measuring segregation required costly, privacy-sensitive mobility data, limiting analysis to regions with high digital penetration. By contrasting this, our framework demonstrates that the built environment itself encodes a legible ``visual grammar'' of social isolation—observable directly from open satellite and street-level imagery. This methodological advance offers significant academic value: it decouples social sensing from surveillance, enabling the scalable, privacy-preserving measurement of human behavioral patterns across entire nations. By synthesizing pixel-level semantics with sociological theory, we move the field from opaque prediction to interpretable sociophysical measurement, establishing that the physical arrangement of our cities—from zoning textures to infrastructure barriers—serves as a primary, quantifiable determinant of social mixing.

A cornerstone of this work is the successful integration of qualitative sociological theory with large-scale computational observation. By leveraging LLMs to synthesize fragmented domain literature into an interpretable visual codebook, we bridged the historical divide between ``thick'' qualitative descriptions of urban life and ``thin'' quantitative measures. The empirical validation of this codebook -- demonstrated by the strong alignment between theoretical cues (\textit{e.g.}, High Fences) and mobility outcomes -- provides compelling evidence that sociological theories of defensible space''~\cite{newman1973defensible} and new urbanism''~\cite{jacobs1961death, talen1999sense} hold true at the scale of entire metropolitan regions. This alignment validates Henri Lefebvre’s theory of the \textit{social production of space}~\cite{henri1991production}, confirming that visual forms actively regulate socioeconomic behavior. This transparency addresses a critical bottleneck in computational social science: the challenge of auditability. By using chain-of-thought prompting~\cite{wei2022chain} to explicitly link visual evidence to behavioral predictions, our approach offers social scientists and planners a granular explanation of why an area is segregated, enhancing trust and enabling mechanism-driven inquiry.

Beyond theoretical validation, this framework functions as a scalable instrument for auditing urban governance and policy impact. Our analysis of inclusionary housing reveals that policy interventions leave a distinct visual signature: communities with such programs are characterized by mixed-use morphologies and active public spaces that facilitate interaction, rather than just demographic shifts on a spreadsheet. This confirms that effective desegregation policies work by physically reshaping the urban environment to foster connectivity~\cite{wang2023inclusionary}. Consequently, our framework offers policymakers a mechanism-linked toolkit to diagnose segregation drivers ex ante and monitor the on-the-ground impacts of design interventions ex post, utilizing widely available open data.

Looking forward, VISAGE provides a robust blueprint for decoding the physical determinants of diverse human behaviors beyond segregation. Its core contribution—a closed-loop, theory-informed reasoning process—is a generalizable solution for operationalizing complex social concepts at scale. This framework can be readily extended to investigate other critical dimensions of urban life, such as perceived safety, public health, or economic vitality. By bridging high-level social theory with large-scale observation through automated reasoning, VISAGE offers a powerful template for a new era of computational human behaviour research, where the physical foundations of social life are measured, modeled, and understood with unprecedented precision and breadth.

Despite its advancements, this study has limitations. First, while we identify strong associations between the built environment and segregation, this cross-sectional analysis does not establish causal directionality; future work should leverage natural experiments or panel designs to strengthen causal inference~\cite{yabe2025behaviour}. Second, as with any LMM-based approach, our framework may reflect biases present in pretraining data or the academic literature used to construct the codebook, necessitating ongoing auditing and sensitivity analyses. Third, although the visual grammar performs robustly across U.S. cities, transfer to markedly different global urban morphologies (e.g., informal settlements in the Global South) would require localized adaptation of the theoretical codebook.
%TC:ignore

\section*{Methods}

\subsection*{M1 Datasets}

\subsubsection*{Urban imagery data}

We compiled a comprehensive dataset of paired street-level and overhead imagery for 10,030 U.S. communities (census tracts) across 31 metropolitan areas. Street-level images were retrieved from Google Maps Street View and standardized to $512\times512$ pixels. Satellite imagery was sourced from Esri USA Satellite (2019) as $256\times 256$ pixel tiles. Community boundaries were defined using U.S. Census Bureau TIGER/Line tract polygons. For each tract, we collected 40 distinct street-view frames sampled from unique panorama locations within the polygon to maximize visual coverage. Simultaneously, we sampled 7 geographic centers per tract to fetch corresponding satellite tiles at zoom level 17 (ground footprint $\approx 300\times 300$ meters per tile). All imagery was mapped to tract IDs, de-duplicated by location (and heading for street view), and resized. The final dataset provides, per community, a paired set of 40 street-view images and 7 satellite tiles, serving as the multi-view visual basis for downstream sociophysical reasoning.

\subsubsection*{Socioeconomic exposure segregation}

Following formulations in ref.~\cite{moro2021mobility,yabe2023behavioral}, we define socioeconomic exposure segregation at community $c$ as the deviation of the income composition of visitors to $c$ from an even, well-mixed distribution. Let $\tau_{c,q}$ denote the share of all visits to $c$ made by residents in the $q$-th income quartile ($q\in\{1,2,3,4\}$). The segregation index $S_c$ is calculated as:

\begin{equation} \label{equ:seg} S_c=\frac{2}{3}\sum_{q=1}^4\left|\tau_{c,q}-\frac{1}{4}\right|, \end{equation}

where $S_c=0$ indicates perfectly even exposure across groups and $S_c=1$ indicates visits are concentrated entirely within a single group.

Ground-truth visitor flows were derived from SafeGraph Weekly Patterns over a 12-week period (2019-08-12 to 2019-11-03), which aggregates anonymized visits to points of interest (POIs). For each metropolitan area, we first ranked communities by ACS 2019 median household income to define the four city-specific income quartiles. Within every target community, we aggregated all visits to POIs inside its boundary, preserving the visitor's home census tract. Each visit was assigned to an income quartile based on the origin tract’s rank. We then computed the visitor shares $\tau_{c,q}$ and the resulting index $S_c$. This procedure yielded a single, mobility-derived segregation score for each of the 10,030 communities, serving as the ground truth for validation.

\subsection*{M2 The VISAGE framework}

As illustrated in Figure~\ref{fig:framework}, VISAGE operationalizes a closed-loop, knowledge-integrated urban sensing paradigm. Rather than learning opaque correlations, the framework integrates multi-view urban imagery with sociological theory to reason about exposure segregation. The workflow consists of three automated stages: (i) Theoretical synthesis, where LLMs distill domain literature into a visual codebook; (ii) Operationalization, where theoretical cues are converted into machine-readable reasoning steps; and (iii) Perception and validation, where a domain-specific LMM performs interpretable prediction and triggers an iterative refinement loop.

\subsubsection*{Theoretical synthesis and codebook generation}

To ground the model in established social science, a multi-LLM Literature Agent module assembles a cross-domain corpus and distills it into a structured visual codebook. We queried academic databases using Boolean combinations of terms related to the built environment, urban form, and segregation, yielding $N=44$ full-text items after de-duplication and screening (Supplementary Table S1).

Three specialized LLM modules executed the distillation. The \textit{Curator} first handled retrieval and metadata normalization (see Supplementary Note S1 for query templates). Next, the \textit{Extractor} parsed full texts to identify image-observable elements and their reported associations with segregation across sociology, urban planning, transportation, and environmental science. For each element, the agent assigned an imagery modality, a provisional category label, an expected direction of association (enhance or alleviate segregation), and a brief rationale. Finally, the \textit{Coder} consolidated these outputs by harmonizing terminology and mapping features to a fixed domain vocabulary. The resulting codebook comprises $F_{sv}=29$ street-view cues and $F_{rs}=15$ remote-sensing cues (Supplementary Tables S2 and S3). Each entry is machine-readable—containing name, definition, modality, and expected direction—providing the prior hypotheses necessary for the downstream reasoning process.

\subsubsection*{Operationalizing theory into visual reasoning}

To convert abstract theoretical cues into observable evidence, the Experiment Agent transforms the codebook and raw imagery into structured supervision. Inputs include the paired imagery for each community and the modality-specific codebooks. Using a fixed prompting scheme, a multi-modal detector~\cite{bai2025qwen2} inspects each image to generate a binary presence vector corresponding to the codebook cues.

For a community $c$, detections are aggregated to form a frequency table $n_c^m\in \mathbb{N}^{F_m}$ per modality $m\in\{SV, RS\}$. Normalized frequencies $\hat{n}_c^m$ serve as a compact, domain-labeled summary of the built environment. The agent then auto-generates codebook-aware Chain-of-Thought (CoT) traces using a four-step template (Supplementary Figure S4): (1) \textbf{Summary} states the task; (2) \textbf{Caption} interprets detected cues using codebook directions; (3) \textbf{Calculation} composes cues into estimated income shares; and (4) \textbf{Answer} reports the final indices. These traces anchor the supervision in observable physical evidence, enforcing multi-cue composition over single-feature shortcuts.

\subsubsection*{Mechanism-aware perception and iterative refinement}The core predictive engine is the Perception Agent, a domain-adapted, cross-view LMM~\cite{zhang2025urbanmllm} that jointly processes the paired imagery and the serialized cue frequencies. The model ingests the visual data alongside a structured prompt encoding the codebook-aware QA traces. A lightweight regression head maps the fused representation to the visitor income composition $\hat{p}_c$ and segregation index $\hat{S}_c$. Training utilizes a multi-task objective aligning predictions with mobility-derived ground truth, while the stepwise reasoning traces provide intermediate supervision to ensure interpretability.

Crucially, the workflow employs an automated feedback loop to validate and refine the theoretical codebook. Following out-of-sample evaluation, we perform stability checks to enforce sign-consistency. If a cue repeatedly violates its theoretically expected direction across folds, a pruning test is triggered: cues are removed or regularized if their exclusion does not statistically degrade the out-of-sample correlation $r$. This ensures the codebook remains empirically supported. Additionally, the system analyzes reasoning traces for compositional inconsistencies, generating corrective meta-instructions to refine the Experiment Agent's prompts. Finally, city-level error maps identify anomalies; the Perception Agent summarizes uncodified visual features in high-error regions, seeding targeted literature queries for the Literature Agent to discover novel cues for the next iteration.

\subsection*{M3 Implementation details}

\subsubsection*{Baseline models}

To validate the contribution of our theory-integrated approach, we compared VISAGE against five baselines (Supplementary Figure S7) spanning three levels of information granularity: semantic structure, language descriptions, and generic visual embeddings.

\begin{itemize}[leftmargin=*]
    \item \textbf{SEG-REG (Semantic Structure):} Street-view images are processed with a scene-parsing model (ResNet18-dilated + PPM\_deepsup on ADE20K) to obtain pixel shares for 150 classes. The top 40 frequent elements are averaged into a community-level descriptor and fed to standard regressors (ElasticNet, XGBoost, Random Forest, SVR, KNN).

    \item \textbf{DESC-REG (Language Descriptions):} To capture finer semantics (e.g., architectural style), we use Qwen2.5-VL-7B to generate visual-element count tables for both street and satellite views. These are merged into a 44-D vector and regressed onto $S$ using the same family of regressors.

    \item \textbf{EMB-REG (Generic Embeddings):} Images are encoded with a pretrained ResNet-50 (ImageNet). 2048-D embeddings are averaged and reduced via PCA to 30 principal components before regression.
    
    \item \textbf{Open-source VLM:} We adopt ViLA~\cite{lin2024vila} as a model-level baseline. Pooled embeddings are extracted and mapped to $S$ via a regression head (MSE objective), isolating VLM feature quality under the same supervision without domain adaptation.
    
    \item \textbf{VISAGE w/o CoT (Ablation):} This variant retains the VISAGE backbone and training target but replaces the codebook-anchored CoT prompts with generic prompts (e.g., “consider housing, infrastructure...”). This isolates the specific contribution of the structured, theory-grounded reasoning process.

\end{itemize}

\subsubsection*{Experiment settings}

\begin{itemize}[leftmargin=*]
\item \textbf{Data splits.} We consider 31 U.S. metropolitan areas comprising 10,030 communities in total. A fixed hold-out of 2,006 communities is reserved as an independent test set. The remaining 8,024 communities form the development pool used for model selection and ablations. Within this pool, we apply 5-fold cross-validation with community-disjoint folds: in each fold, 80\% of communities are used for training and 20\% for validation. All baselines and ablated variants use the same splits. Each community provides 7 satellite tiles and 40 street-view images.

\item \textbf{Training details.} UrbanMLLM’s visual encoders are initialized from SigLIP pretrained weights. Unless otherwise specified, optimization uses AdamW with a cosine learning-rate schedule (initial LR 1×10$^{-5}$), batch size = 2, and early stopping on validation loss. Random seeds are fixed across folds for reproducibility. For regression-based baselines (SEG-REG, DESC-REG, EMB-REG), we train ElasticNet, XGBoost, Random Forest, SVR, and KNN under identical search spaces and report the best validation configuration per fold.

\item \textbf{Feature post-processing.} When a method produces a fused embedding from the cross-view module, we extract a 2048-dimension vector and apply PCA to 30 dimensions before downstream regression; this setting yielded the most robust validation performance across folds and is kept fixed for all embedding-regression variants, including the VISAGE (w/o CoT) ablation.

\item \textbf{Compute environment.} Experiments run on 4 NVIDIA A100 GPUs; core software includes PyTorch 2.3.0, CUDA 12.2.

\item \textbf{Evaluation metrics.} Performance is assessed on the held-out test communities using coefficient of determination ($R^2$), mean absolute error (MAE), and mean squared error (MSE). Metrics are computed per fold and reported as the mean across folds.

\end{itemize} 

\subsection*{M4 Policy impact analysis}

To quantify the physical impact of policy, Inclusionary Housing (IH) locations were compiled from the Inclusionary Housing Map \& Program Database and spatially matched to community polygons ($IH_c=1$ if present). We estimated cross-sectional OLS models with city fixed effects to compare communities with and without IH, adjusting for demographic covariates $D_c$ (ethnic composition and socioeconomic status):

$$S=\alpha+\beta IH_c+\delta^TD_c+\mu_{city}+\epsilon_{c}.$$

$$\hat{S}=\hat{\alpha}+\hat{\beta} IH_c+\hat{\delta}^TD_c+\hat{\mu}_{city}+\hat{\epsilon}_{c}.$$

Here $\mu_{city}$ denotes city fixed effects. Standard errors were clustered by metropolitan area. The coefficients $\beta$ and $\hat{\beta}$ capture the average difference in ground-truth and predicted segregation, respectively, associated with IH policies after controlling for local demographics and city-level heterogeneity.

\renewcommand*{\thefigure}{S\arabic{figure}}
\renewcommand*{\thetable}{S\arabic{table}}
\renewcommand*{\thesection}{S\arabic{section}}
\setcounter{figure}{0}
\setcounter{table}{0}
\section*{Supplementary Notes}

\section{The VISAGE framework}

VISAGE is a knowledge-integrated, multi-modal reasoning system designed to measure socioeconomic exposure segregation using only open imagery from street-view and satellite sources. Rather than learning opaque, black-box correlations between pixels and labels, the framework builds an interpretable vocabulary of visual cues synthesized from domain literature. It then utilizes this vocabulary to reason from scene semantics to visitor income composition and a segregation index. The framework executes a closed-loop scientific discovery process powered by three main functional modules: the Literature Agent (which distills theory into a visual codebook), the Experiment Agent (which operationalizes cues into structured reasoning templates), and the Perception Agent (which performs mechanism-aware cue detection and inference).

\subsubsection*{Literature agents: Theoretical synthesis}

This module transforms diffuse domain knowledge into two interpretable, machine-readable codebooks of image-observable cues—one for street-view observables and one for remote-sensing observables. The Literature Agent implements a rigorous three-stage workflow (\textit{Curator}, \textit{Extractor}, \textit{Coder}) to separate concerns, enforce strict input–output schemas, and ensure auditability.

\textbf{Curator}. The Curator performs multi-database retrieval using predefined Boolean queries that combine terms for the built environment and urban form with exposure segregation and social mixing. Screening is conducted based on title and abstract, strictly requiring an empirical link between image-observable built-environment elements and segregation or mixing. Mobility-only studies without image-mappable cues are excluded. After de-duplication (by DOI or title) and metadata normalization (authors, venue, year), the Curator ranks items by relevance and citation count. It outputs two artifacts: PRISMA flow counts and a machine-readable table of accepted items, including perspective hints for downstream parsing. The curated corpus comprises 44 studies (13 sociology, 10 urban planning, 8 housing, 6 transportation, 7 environment), listed in Supplementary Table~\ref{tab:S1}. The workflow is illustrated in Supplementary Figure~\ref{fig:S1_1}.
\textbf{Extractor}. The Extractor ingests full texts from the Curator list and parses Methods, Results, and Discussion sections to identify cues that are observable in images and explicitly linked to segregation or mixing. For each cue it records a raw name, a provisional imagery perspective (street view, remote sensing, both, or unclear), a provisional family label, the reported direction of association (increase segregation, decrease segregation, or ambiguous), a short distilled rationale, and one short verbatim evidence quote with a page or figure anchor. Optional appearance hints are recorded when the paper describes how the cue typically presents in images. The Extractor produces a structured list of candidate cues that is source-anchored and ready for consolidation. The workflow is illustrated in Supplementary Figure~\ref{fig:S1_2}.

\textbf{Coder}. The Coder consolidates Extractor outputs by normalizing terminology, merging near-duplicates using string-similarity thresholds, and mapping cues to a fixed family ontology. Final imagery perspective is assigned using a rubric that distinguishes street-scale façades, signage, sidewalks, and fence transparency from overhead roof, parcel, block, and network patterns. For each cue the Coder provides a concise definition and detection hints. Remote-sensing entries include a short ``visual manifestation'' that describes overhead appearance. Reported directions are combined by confidence-weighted voting to yield a direction consensus (increase, decrease, or ambiguous). Unresolved contradictions are listed in a conflict register. The output consists of two codebooks, a remote-sensing (satellite) codebook and a street-view codebook, provided in Supplementary Table~\ref{tab:S2} and~\ref{tab:S3} respectively, plus a small conflict file for audit. We adopt ten families to structure reasoning later in the pipeline: access and permeability; housing diversity and type; commercial frontage; maintenance and vacancy; greenery and shade; road structure; barriers and boundaries; public space; transport hubs; and land-use mix. The finalized codebooks contain 29 street-view cues and 15 remote-sensing cues. Quality control includes field-completeness checks, random reviewer audits with agreement statistics, versioning of the codebook, and checksums for all artifacts. The workflow is illustrated in Supplementary Figure~\ref{fig:S1_3}.

\subsubsection*{Experiment agent}

\begin{figure}[htbp]
\centering
\includegraphics[width=1.0\linewidth]{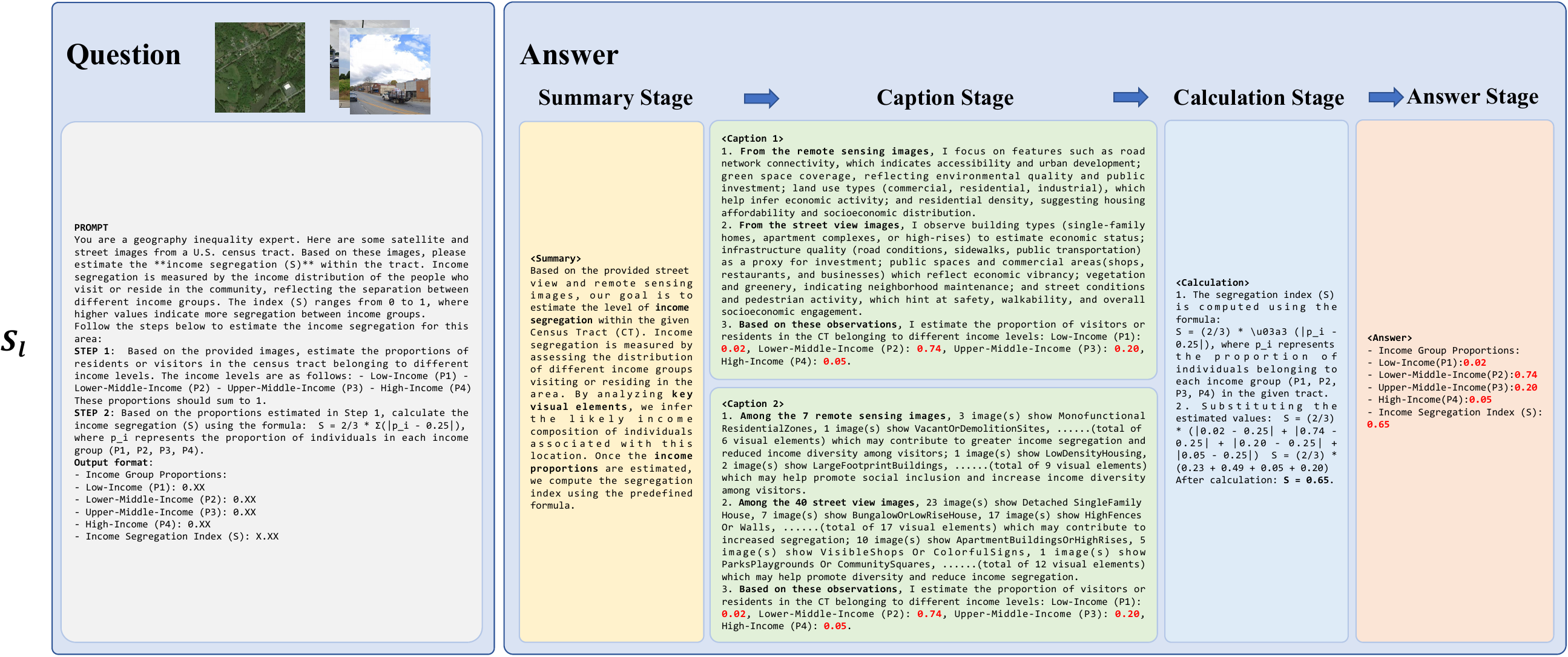}
\caption{An example of the entire Chain-of-Thought prompt design for the Question-Answer pair.} 
\label{fig:S2}
\end{figure}

This agent converts the extracted cue evidence into a structured reasoning trace, which guides the subsequent inference by the Perception Agent. Each inference round consists of a single Question (the prompt to the model) and a four-stage Answer. The Question binds inputs and constraints; the Answer exposes the model's reasoning in labeled stages so that predictions can be audited and replicated (Supplementary Figure~\ref{fig:S2}).

\textbf{Imagery alignment and cue detection}. For each community, we first collect paired street-view and satellite tiles that are geospatially aligned to census tracts. The Perception Agent performs cue detection guided by the two codebooks produced by the Literature Agents. Per-image detections are aggregated to community-level frequency tables for each modality. We also store normalized frequencies so that later stages can use compact summaries of the built environment without re-running detection.

\textbf{Question.} The Question template, generated by the Experiment agent, presents only information available from open data and the codebook. It has four parts:

\begin{itemize}[leftmargin=*]
    \item \textbf{Task statement}. Define the goal: estimate a community's visitor income composition and the exposure-segregation index from street-view and satellite imagery and the derived cue frequency tables.
    \item \textbf{Inputs}. Imagery context: identifiers for the set of tiles (street view and satellite). Cue evidence: the community's codebook-derived cue frequency tables for each modality (normalized counts). City context: the four income groups are city-specific quartiles.
    \item \textbf{Reasoning requirements}. Consider multiple cue families jointly (e.g., access/permeability; housing diversity/type; commercial frontage; maintenance/vacancy; greenery/shade; road/network structure; barriers/boundaries; public space; transport hubs; land-use mix). Prefer explanations that refer to families rather than single cues. Do not use external data or private information.
    \item \textbf{Output contract.} Require the model to return a labeled, four-stage Answer; numeric outputs must include the four income shares (sum to 1) and the scalar index. Require all intermediate numbers used in the final calculation to be shown.
\end{itemize}

\textbf{Answer}. The model must structure its response in four stages, each labeled in the text and emitted in order.

\begin{itemize}[leftmargin=*]
    \item \textbf{Summary}. Restate the task in one or two sentences, cite that only open imagery and codebook cues are used, and list the cue families that will be considered jointly for this community. No numbers appear in this stage.
    \item \textbf{Caption}. Interpret the community's cue frequency tables by grouping salient cues under the families above. For each family, write short, literature-aligned sentences linking high or low prevalence to expected mixing or separation. Examples: higher permeability, mixed land use, and public space openness are associated with more cross-group encounters; high fences, cul-de-sacs, and monofunctional residential blocks are associated with reduced encounters. The Caption must reference multiple families and avoid single-cue explanations.
    \item \textbf{Calculation}. Translate the Caption into a visitor income composition (four quartiles defined within the city) and compute the exposure-segregation index. All intermediate values are written explicitly before the final index. Monotonicity guidance is enforced (\textbf{e.g.}, more barriers should not decrease the index; more diversity should not increase it, unless clearly justified).
    \item \textbf{Answer}. Report the four income shares and the final index, followed by a brief rationale that names the most influential families (not individual cues) and ties back to the Caption. No new assumptions may be introduced here.
\end{itemize}

\subsubsection*{Perception agent}

This agent, implemented as a domain-adapted Large Multi-modal Model (LMM), performs the visual cue detection, executes the structured reasoning, and infers the final segregation index. This module aligns predictions with ground truth while preserving interpretable intermediate evidence.

\textbf{Inputs and targets}. Each training instance includes the set of street-view and remote-sensing images for a community and the serialized cue frequency tables produced in the previous stage. Targets are the visitor income composition and the exposure-segregation index computed from mobility data for the year 2019.

\textbf{Architecture.} Images are encoded and fused by a cross-view component to produce a multi-view embedding. The cue tables are embedded by a small projection network and concatenated with the visual embedding. A lightweight regression head maps the joint embedding to a predicted segregation index. In a regularization variant, we add an auxiliary head to predict the income composition before the index.

\textbf{Objective and protocol}. Unless otherwise stated, training minimizes mean squared error between the predicted and true segregation index. See detailed experimental settings in the following sections.

\subsubsection*{Closed-Loop Scientific Discovery}

The agents described above operate within a closed-loop scientific discovery process. This framework is dynamic, allowing for continuous, data-driven refinement of the theoretical knowledge base and reasoning protocols. The loop comprises three distinct, self-correcting feedback mechanisms driven by empirical data:

\textbf{Hypothesis refinement.} We perform continuous stability checks on all codebook cues. If a cue's empirical correlation with the segregation index ($S$) violates its expected theoretical direction (sign-consistency) across a defined threshold of cross-validation folds, we initiate a pruning test. Only if the removal of the cue does not cause a statistically significant degradation in the framework's overall out-of-sample correlation ($r$) is the cue subsequently regularized or removed from the visual codebook. This ensures the knowledge base is efficient and empirically validated.

\textbf{Reasoning Protocol Tuning.} The CoT traces are analyzed to identify compositional inconsistencies—instances where the LMM's internal logic violates the expected influence of the cues (e.g., misinterpreting an open cue as contributing to separation). These inconsistent trajectories are used to generate corrective meta-instructions. This feedback is routed to the Experiment agent to systematically optimize and enhance its prompt templates, ensuring a more robust and theoretically consistent compositional logic in subsequent reasoning signal generation.

\textbf{New Knowledge Generation.} Prediction errors are aggregated at the city level to identify low-reliability urban areas, which are treated as locations of potential un-codified mechanisms. The Perception agent is prompted to summarize salient, un-codified visual anomalies in the high-error images from these cities. This summary becomes the input for a targeted city-specific literature query delivered to the Literature agents, initiating a search for novel, regional theoretical concepts to be formalized as candidate codebook cues for the next iteration.

\section{Implementation details}

\subsection{Datasets}

\begin{figure}[htbp]
\centering
\includegraphics[width=0.9\linewidth]{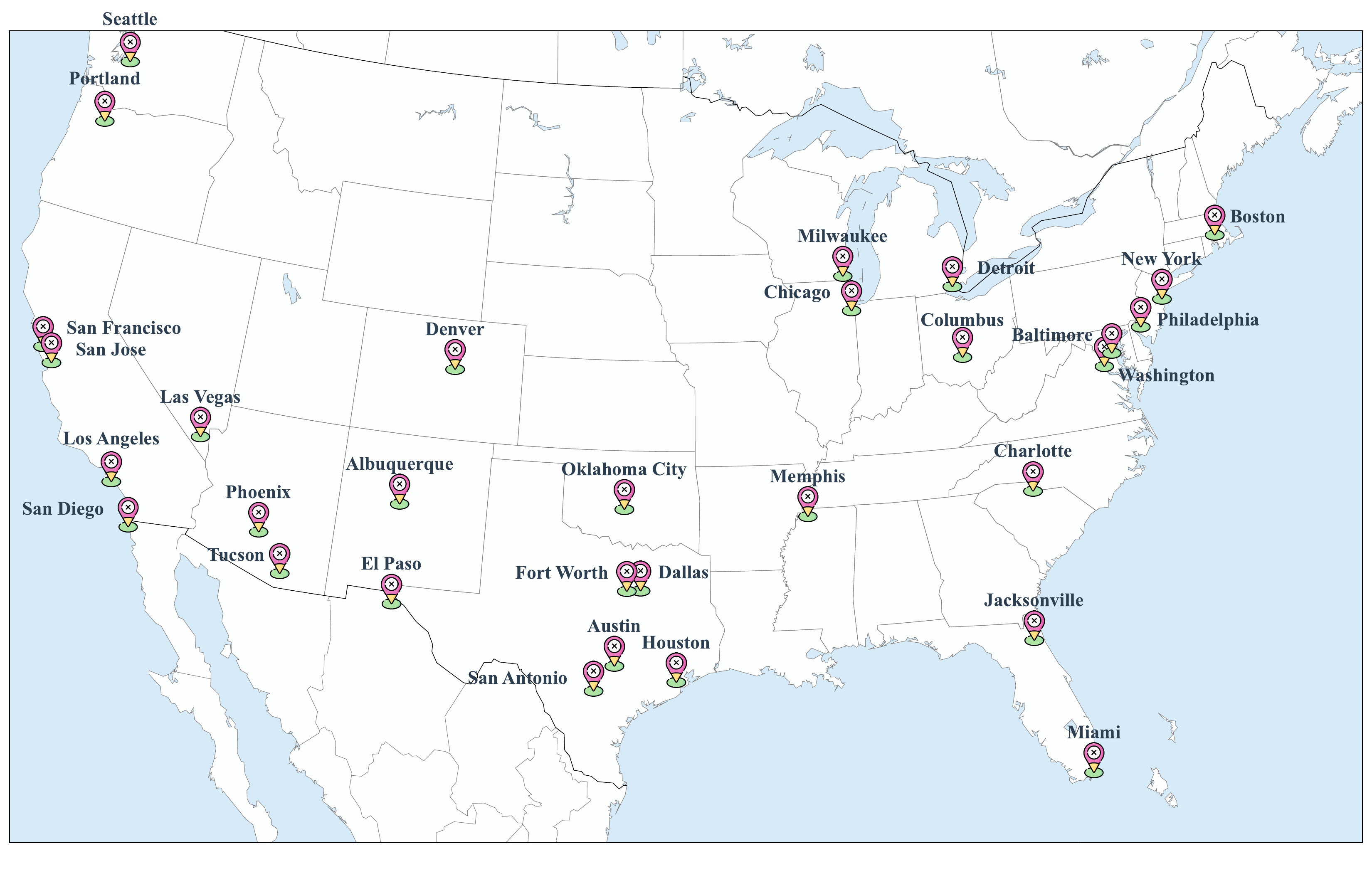}
\caption{The selected 31 most populated cities in the U.S.} 
\label{fig:S3}
\end{figure}

The robustness and generalizability of the VISAGE framework are demonstrated through rigorous evaluation on a comprehensive, multi-modal dataset spanning 31 major U.S. cities (Supplementary Figure~\ref{fig:S3}). The dataset construction focuses on maximizing the visual information input and establishing a granular, empirically-derived ground-truth metric for socioeconomic exposure segregation.

\subsubsection*{Urban imagery data}

The visual input layer relies on a curated collection of paired street-level and aerial imagery, systematically collected to maximize coverage and visual information per community. Street-level imagery ($I_{SV}$) was retrieved from the Google Maps Street View service and uniformly standardized to a resolution of 512×512 pixels. To mitigate directional and spatial sampling bias while ensuring high visual information content, we significantly augmented the typical sampling density. For each Census Tract $l$, we collected up to 40 distinct street-view frames ($I_{SV,l}$) sampled from unique panorama locations entirely within the official geographic boundaries defined by the U.S. Census Bureau TIGER/Line Shapefiles~\cite{Census_TIGERLine_series}.

The satellite imagery ($I_{SI}$) provides critical context on urban form, density, and land use. This data originates from the Esri USA Satellite~\cite{EsriUSASatellite2019} service and is stored as 256×256 pixel tiles. To enhance the effective visual resolution and ensure the imagery is focused on the community’s internal structure, the image sampling was conducted at a high zoom level 17. At this magnification, each tile corresponds to a compact ground footprint of approximately 300×300 meters. Furthermore, instead of relying on a single aerial view, we sample seven distinct geographic center points per Census Tract $l$, resulting in a set of 7 tiles $I_{SI,l}$. This multi-tile sampling strategy ensures that the collective aerial views comprehensively cover the spatial extent of the average-sized Census Tract. All collected images were systematically de-duplicated, resized, and precisely mapped to their respective Census Tract ID $l$.

\subsubsection*{Urban mobility data}

The ground-truth label, the socioeconomic exposure segregation index $S$ for each community, is calculated using high-fidelity urban mobility data derived from the SafeGraph Weekly Patterns dataset, which reports anonymized weekly visits to Points of Interest (POIs) and the associated home Census Tracts of visitors. To achieve statistical stability, visit patterns are aggregated across a 12-week time window, spanning from 2019-08-12 to 2019-11-03.

The calculation of the index involves a precise multi-step procedure. First, within each Metropolitan Statistical Area, all Census Tracts are ranked based on their 2019 American Community Survey (ACS) median household income and subsequently partitioned into four equal-sized income groups (quartiles), denoted by $q\in${1,2,3,4}. Second, for a target community $c$, all visits recorded across the 12-week period to all POIs located within its boundary are aggregated. Each visit is then assigned to one of the four income groups $q$ based on the income rank of its origin tract $l$. Third, the metric $S_c$ quantifies the deviation of the observed income composition of visitors, $\tau_{c,q}$ (the share of visits from the $q$-th income quartile), from a state of perfect even mixing, where each quartile contributes exactly 1/4 of the total visits. The segregation index $S_c$ is formally defined as:

$$S_c=\frac{2}{3}\sum_{q=1}^4|\tau_{c,q}-\frac{1}{4}|.$$

In this formulation, $S_c$ ranges from 0 to 1, where $S_c=0$ signifies perfectly even exposure across all four income groups, and $S_c=1$ indicates maximal segregation, with visits entirely concentrated within a single income group. The city-average $S$ and the distribution of $S_c$ are illustrated in Supplementary Figure~\ref{fig:S5}.
  
\subsubsection*{Inclusionary housing}

To provide a rigorous, external validation of the VISAGE framework's sensitivity to real-world policy effects, we incorporated auxiliary data on Inclusionary Housing (IH) policies. These policies typically mandate that developers set aside a fixed percentage of new residential units as affordable housing, representing a direct, governmental intervention aimed at increasing socioeconomic mixing and reducing residential segregation. We compiled a dataset indicating the presence or absence of an active, mandatory IH policy within the jurisdiction of each studied Census Tract during the 2019-2021 period~\cite{GH_InclusionaryHousingMap}. This policy data facilitates an advanced regression analysis (detailed in Supplementary Table~\ref{tab:S4}) to confirm whether the visual cues identified by VISAGE and its predicted segregation index $S_c$ statistically align with the expected policy-induced changes in ground-truth socioeconomic exposure.

\subsection{Baseline models}

To comprehensively benchmark the performance and validate the methodological advancements of the VISAGE framework, we structured our experiments around three distinct model categories: Conventional Computer Vision (CV) baselines, Large Multi-Modal Model (LMM) baselines, and the proposed VISAGE Framework. These categories allow us to systematically compare the performance gains achieved by leveraging deep visual feature extraction, general multi-modal understanding, and finally, our unique theory-grounded reasoning approach. The following paragraphs detail the specific implementation of each model within these three categories.

\subsubsection*{Conventional CV methods}

\begin{figure}[htbp]
\centering
\includegraphics[width=0.8\linewidth]{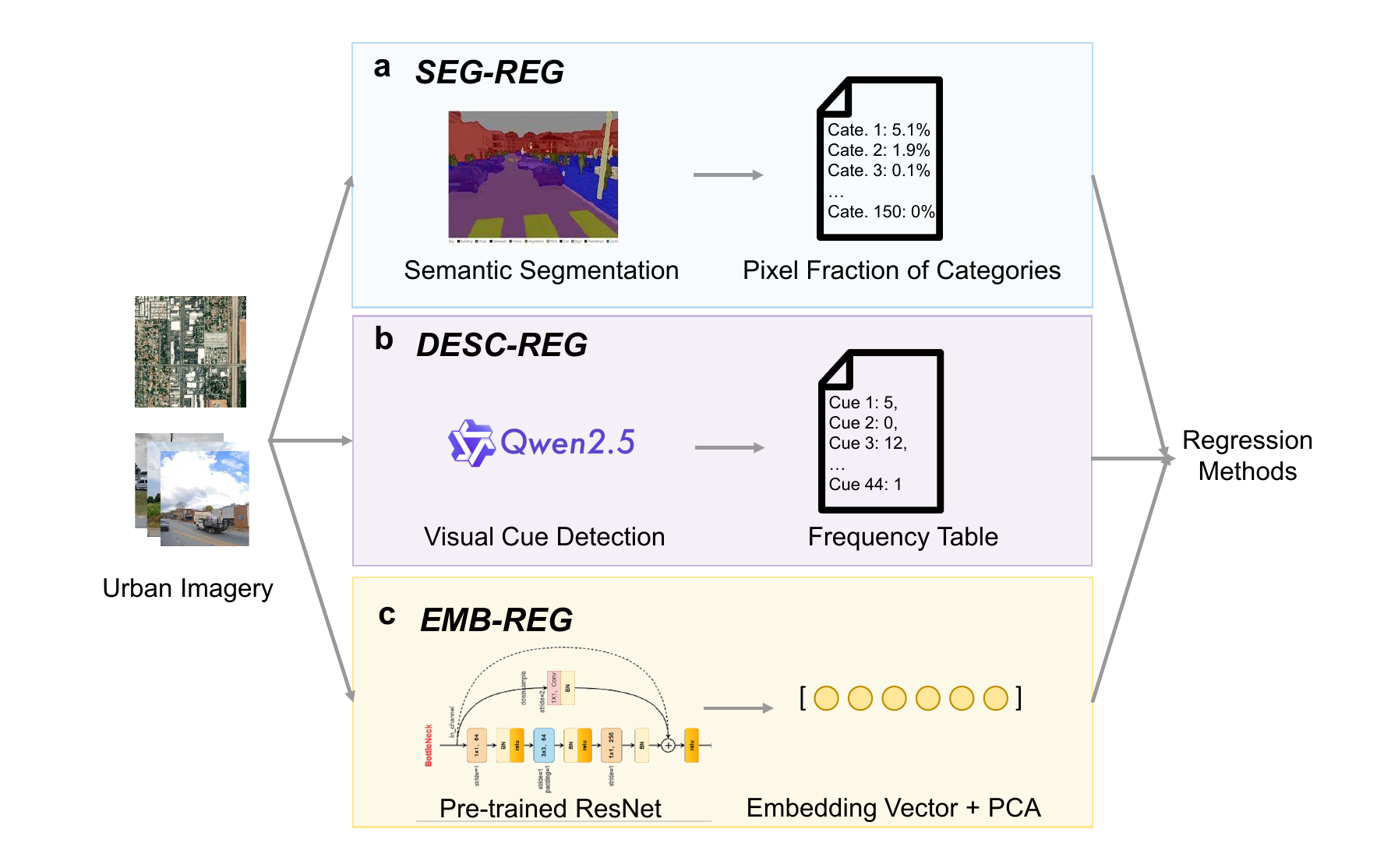}
\caption{Architectural overview of the conventional CV baselines.}
\label{fig:S6}
\end{figure}

Please refer to Supplementary Figure~\ref{fig:S6} for the architectural diagrams of the three conventional CV methods.

\subsubsection*{Semantic segmentation-based regression (SEG-REG)} 
\setlength{\parindent}{0pt}

The core idea of the SEG-REG baseline is to leverage high-performance computer vision models for semantic segmentation, extracting fine-grained urban visual elements (objects and scene categories) from street-view imagery, and subsequently using the density of these elements as features for segregation prediction. This approach assesses the effectiveness of general, low-level visual features in this specific task.

\textbf{Feature Extraction}: We employed a state-of-the-art semantic segmentation model with a ResNet18-dilated architecture paired with PPM\_deepsup (Pyramid Pooling Module with deep supervision), which was pre-trained on the large-scale MIT ADE20K scene parsing dataset~\cite{zhou2018semantic}. This model is capable of classifying 150 different semantic categories and outputting the pixel-wise proportion for each class within an input image.

\textbf{Feature Vector Construction}: For each street-view image, we extracted the pixel proportions for all recognized semantic categories. We then identified the top 40 most frequently occurring semantic elements across the entire dataset to serve as the feature variables for the regression model.

\textbf{Community-Level Aggregation}: To obtain a community-level representation, we aggregated the features across all street-view images belonging to a single community by calculating the mean pixel proportion for each of the 40 visual element categories, resulting in a 40-dimensional feature vector for each community.

\textbf{Regression Modeling}: We utilized a suite of popular and robust machine learning regression algorithms, including ElasticNet, XGBoost, Random Forest, Support Vector Regression (SVR), and K-Nearest Neighbors (KNN), to train and predict the income exposure segregation index $S$ based on this feature vector.

\subsubsection*{Image description-based regression (DESC-REG)}
\setlength{\parindent}{0pt}

While semantic segmentation excels at conventional objects (e.g., roads, buildings, vegetation), it often lacks the capacity to capture more nuanced or socioeconomically relevant visual details (e.g., architectural style, specific land use, quality of green spaces). Furthermore, the SEG-REG model is inherently limited to street-view imagery. The DESC-REG baseline addresses this limitation by exploiting the advanced multi-modal understanding capabilities of LMMs to generate richer, text-based visual features from both satellite and street-view images.

\textbf{Feature Generation}: We utilized the Qwen2.5-VL-7B model~\cite{qwen2.5-VL}, which was also used in our VISAGE framework, to generate comprehensive textual descriptions for both street-view and satellite images. These descriptions were then aggregated into community-level visual element statistics tables, covering more specific and relevant visual cues such as detailed building styles, urban infrastructure types, and green space classifications. These elements are more aligned with the specific visual characteristics known to correlate with socioeconomic segregation.

\textbf{Feature Vector Construction}: The descriptive information from both street-view and satellite images was converted into a unified statistical table. This process yielded a 44-dimensional visual element feature vector for each community, serving as the input for the regression models.

\textbf{Regression Modeling}: Similar to the SEG-REG baseline, we employed the same range of multivariate regression algorithms (ElasticNet, XGBoost, Random Forest, SVR, and KNN) for training and evaluating the prediction of the segregation index $S$ on the test set.

\subsubsection*{Image embedding-based regression (EMB-REG)}
\setlength{\parindent}{0pt}

Both SEG-REG and DESC-REG rely on pre-defined or extracted visual element categories, which inevitably leads to the loss of potential implicit or holistic information contained within the raw images. The EMB-REG baseline is designed to overcome this by employing a more generic and dense image representation, aiming to capture the overall semantic context of the urban scenes.

\textbf{Image Embedding Extraction}: We first utilized the pre-trained ResNet-50 model~\cite{he2016deep}, trained on the extensive ImageNet dataset, to extract a global, deep visual feature vector (embedding) for every individual street-view and satellite image. The ResNet-50 output provides a 2048-dimensional image embedding that encapsulates high-level semantic features.

\textbf{Community-Level Aggregation}: To represent a community, the embedding vectors from all associated street-view and satellite images were averaged, generating a single, aggregated 2048-dimensional feature vector.

\textbf{Dimensionality Reduction}: To mitigate the curse of dimensionality and remove potential noise, we applied Principal Component Analysis (PCA) to the aggregated 2048-dimensional vector, retaining the top 30 principal components. This 30-dimensional vector constituted the final, dense visual feature input for each community.

\textbf{Regression Modeling}: The same set of multivariate regression models (ElasticNet, XGBoost, Random Forest, SVR, and KNN) were used to train and assess the predictive performance of the dense feature vector on the income exposure segregation index $S$.

\subsubsection*{Open-source LMM methods}
\setlength{\parindent}{0pt}

These baselines, employing the well-regarded VILA family of models~\cite{lin2024vila}, serve as a control group to measure the performance of a LMM on this task without any domain-specific fine-tuning on our codebook data.

\textbf{VILA without CoT (VILA w/o CoT)}: This model operates in a zero-shot manner, provided only with the image pair and a direct instruction to predict the final segregation index $S$. It tests the model's raw capability to map visual input directly to the numeric outcome, bypassing any intermediate reasoning or cue identification.

\textbf{VILA with CoT (VILA w/ CoT)}: This model uses the same generic VILA but is instructed to generate an explicit step-by-step rationale before providing the prediction of $S$. This comparison with VILA w/o CoT isolates the effect of general-purpose LMM reasoning—how much improvement CoT provides when the LMM relies only on its pre-trained general knowledge, rather than domain theory.

\subsubsection*{VISAGE}

Here, our domain-Aapted LMM without CoT is the most critical ablation, designed to isolate the independent contribution of the structured CoT reasoning process. This baseline utilizes the identical LMM that underpins the full VISAGE framework, meaning it has been fine-tuned and domain-adapted using the data generated by the three Agents (incorporating the theory-grounded visual codebook). However, it is prompted for the direct prediction of $S$ without the instruction to output the step-by-step CoT rationale.

\subsection{Experiment settings}

This section provides comprehensive details on the experimental setup, training hyperparameter choices, architectural configurations, and evaluation methodology necessary for the full reproducibility of the VISAGE framework and all comparative baselines.

\subsubsection*{Evaluation metrics}

Performance is rigorously assessed on the independent, held-out test set using four standard regression metrics. All metrics are computed per fold and reported as the mean across the 5 folds. Please refer to Supplementary Figure~\ref{fig:S7} for the metrics of all methods.

\begin{itemize}[leftmargin=*]
    \item Pearson Correlation Coefficient ($r$): Measures the linear correlation between the model's prediction $\hat{S}$ and the ground truth $S$. This is a crucial metric for establishing the reliability of visual features in sensing macro-level socioeconomic trends.

    \item Coefficient of Determination ($R^2$): Represents the proportion of the variance in the dependent variable $S$ that is predictable from the model.

    \item Mean Absolute Error (MAE): Measures the average magnitude of the errors, providing a robust, interpretable measure of the average prediction error in the original units of the segregation index $S$.

    \item Mean Squared Error (MSE): Measures the average squared difference between predictions and ground truth, penalizing larger errors more heavily.

\end{itemize}

\subsubsection*{Training details}

The VISAGE framework and its LMM ablations are implemented using the UrbanMLLM architecture~\cite{zhang2024housing}. The visual encoders were initialized from SigLIP pretrained weights~\cite{zhai2023siglip}. During the instruction fine-tuning phase, the AdamW optimizer~\cite{loshchilov2019decoupled} was utilized, paired with a cosine learning-rate schedule initiated at $1\times10^{-5}$. Due to the high memory demands associated with processing 47 images per community alongside a large LMM, training was conducted with a small batch size of 2, distributed across 4 NVIDIA A100 GPUs equipped with 80GB VRAM. An early stopping mechanism, based on monitoring the validation loss, was implemented to prevent overfitting, with training capped at a maximum of 5 epochs.

\subsubsection*{Feature post-processing and regression modeling}

In methods that rely on the final fused embedding for prediction (\textit{e.g.}, the VISAGE w/o CoT ablation), the intermediate image-text fusion representation is extracted from the cross-view attention module as a 2048-dimensional vector. To enhance model generalization and reduce computational load, we applied Principal Component Analysis (PCA). Based on comprehensive validation analysis, the features were consistently reduced to 30 dimensions for the final regression training, as this setting yielded the most robust prediction effect across all folds.

For all baselines and the VISAGE ablations, a multi-model comparison approach was used, searching across ElasticNet, XGBoost, Random Forest, SVR, and KNN regressors. The model yielding the best validation result for each feature set was selected for final reporting:

\begin{itemize}[leftmargin=*]
    \item For the full VISAGE framework and the VISAGE w/o CoT ablation (which uses the PCA-reduced fused embedding), Support Vector Regression (SVR) was chosen as the final head.
    \item For the SEG-REG baseline (using semantic segmentation features), Support Vector Regression (SVR) was also chosen.
    \item For the EMB-REG baseline (using ResNet-50 features), the best validation performance was achieved using the K-Nearest Neighbors (KNN) model.
    \item For the DESC-REG baseline (using vectorized text features), the XGBoost Regressor was ultimately selected.
\end{itemize}

For all traditional CV baselines, the optimization was handled using the Adam optimizer~\cite{kingma2014adam} with an initial learning rate of 10$^{-4}$, and a batch size of 64.

\subsection{Performance comparison}

VISAGE's image-based estimates align closely with the mobility-derived ground truth, demonstrating high predictive reliability in inferring exposure segregation across the 31 U.S. cities and 10,030 communities studied. The framework achieves a strong Pearson correlation ($r=0.770$), firmly establishing that the visual landscape of urban environments contains robust, scalable information to perceive complex segregation patterns. This high-fidelity perception is not localized but generalizes consistently across diverse urban contexts, confirming the utility of our approach for fine-grained spatial analysis where mobility data is scarce.

\begin{figure}[htbp]
\centering
\includegraphics[width=0.9\linewidth]{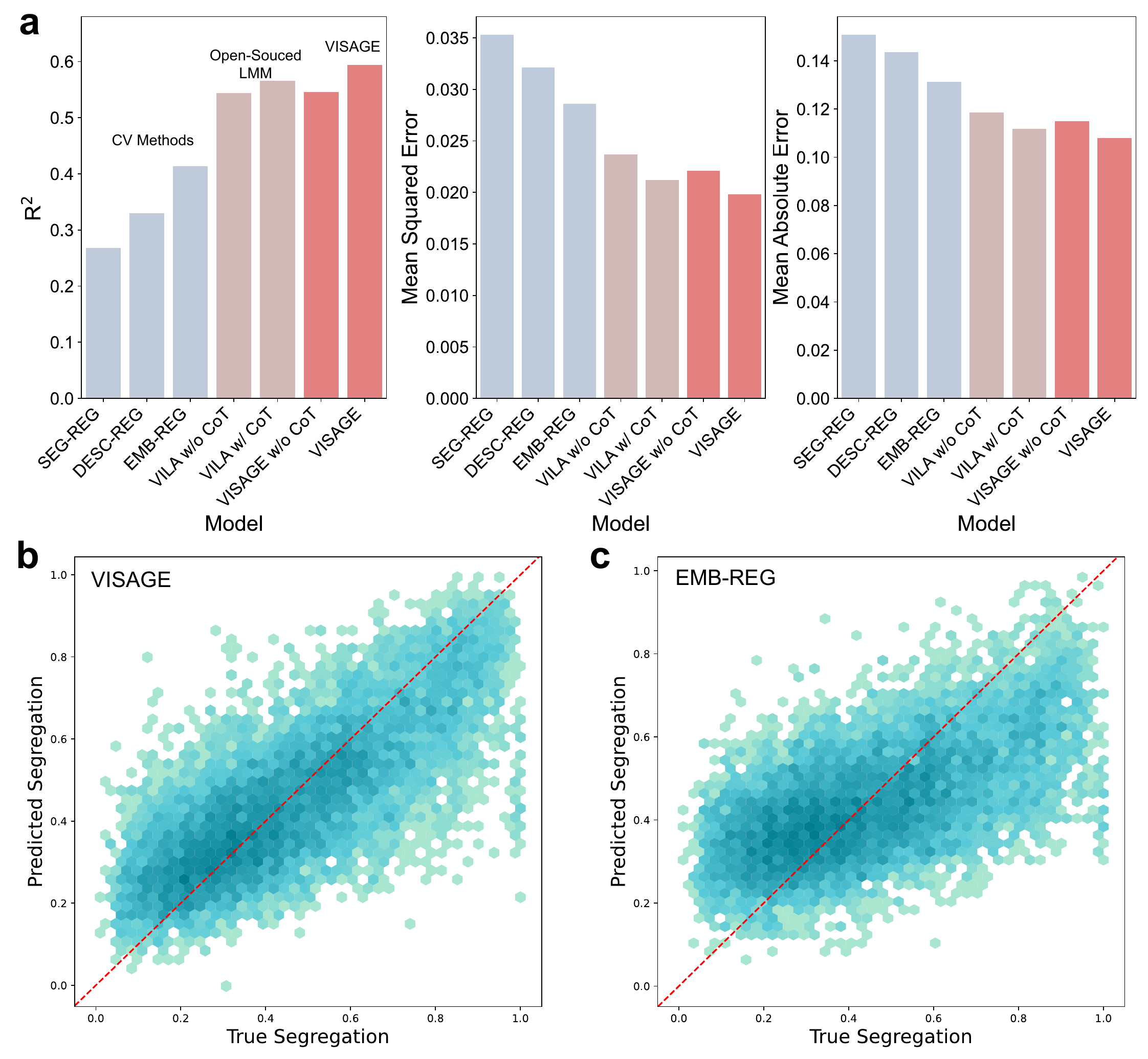}
\caption{\textbf{The performance evaluation of VISAGE.} \textbf{a}, Comparative performance on conventional CV methods, open-source LMM, and VISAGE. Metrics include $R^2$, MSE, and MAE. \textbf{b-c}, Model predictions given by VISAGE (\textbf{b}) and the best-performing conventional CV baseline (\textbf{c}) versus the ground-truth segregation indices.}
\label{fig:S7}
\end{figure}

To validate the methodological contribution of our mechanism-driven approach, we benchmarked VISAGE against a suite of conventional CV models and open-sourced LMMs. VISAGE consistently and significantly outperforms all established correlation-based baselines across all evaluation metrics, showcasing its superior visual reasoning capabilities (Supplementary Figure~\ref{fig:S7}c-d). The visual difference is stark: while correlation-based baselines yield predictions that deviate broadly from the ground truth identity line, VISAGE's estimates are tightly concentrated around the 45-degree line (Supplementary Figure~\ref{fig:S7}a-b). This enhanced fidelity underscores the framework's ability to accurately capture the true magnitude and variation of segregation.

Crucially, this superior performance is directly attributable to the system's knowledge grounding and structured reasoning. Ablation studies confirm that the predictive power is not merely a consequence of model size, but a result of the agentic workflow. Specifically, performance significantly degrades when using ungrounded open-source LMMs or when removing the Chain-of-Thought (CoT) reasoning step. These comparative findings establish that achieving robust and generalizable urban perception requires a methodological shift: moving beyond simple correlational prediction to implement knowledge-integrated, structured reasoning is the material contributor to both reliability and generalization.
\section*{Supplementary Figures}
\addcontentsline{toc}{section}{Supplementary Figures}

\begin{figure}[htbp]
\centering
\includegraphics[width=0.8\linewidth]{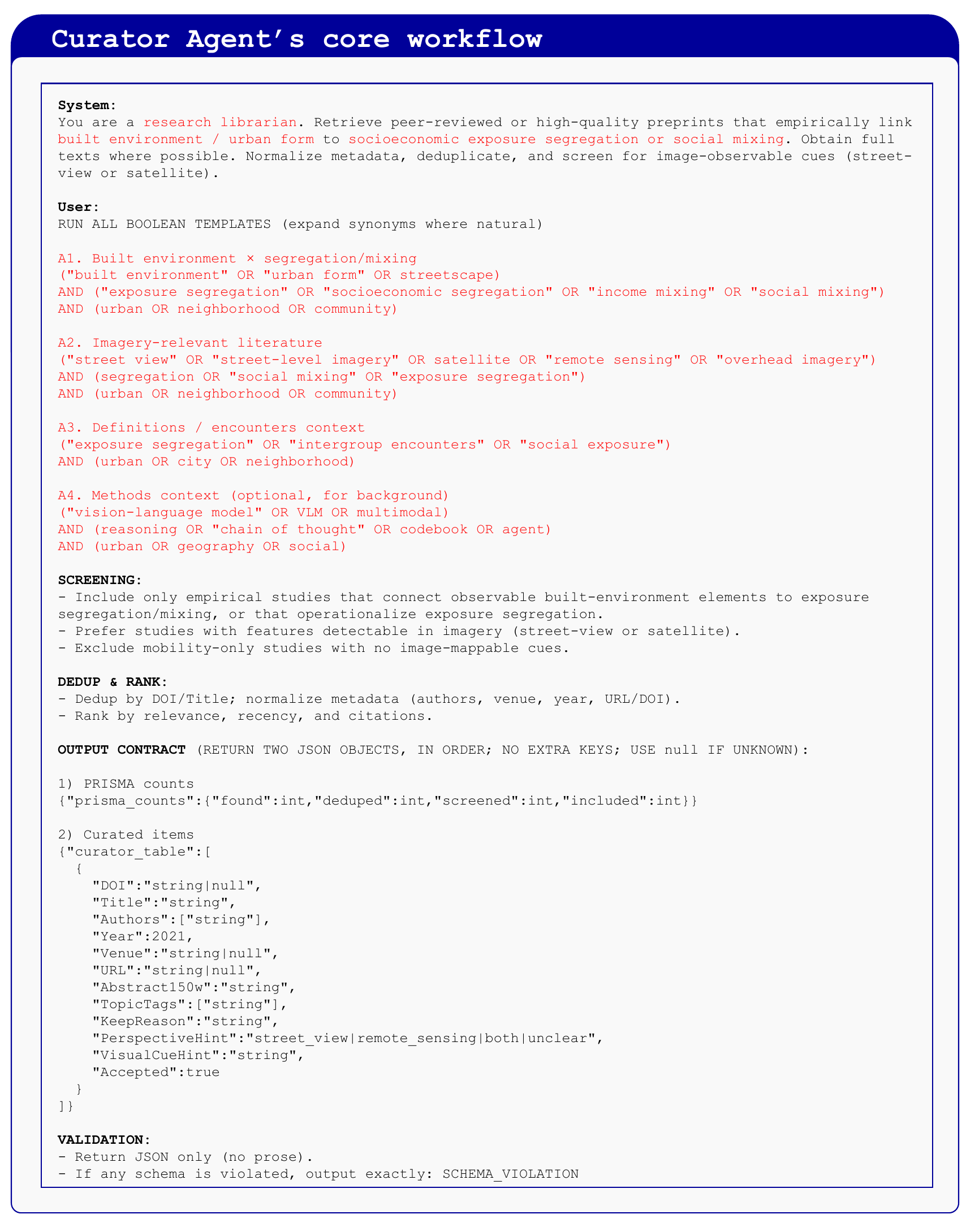}
\caption{Illustration of the agentic workflow of the Curator agent.} 
\label{fig:S1_1}
\end{figure}

\begin{figure}[htbp]
\centering
\includegraphics[width=0.8\linewidth]{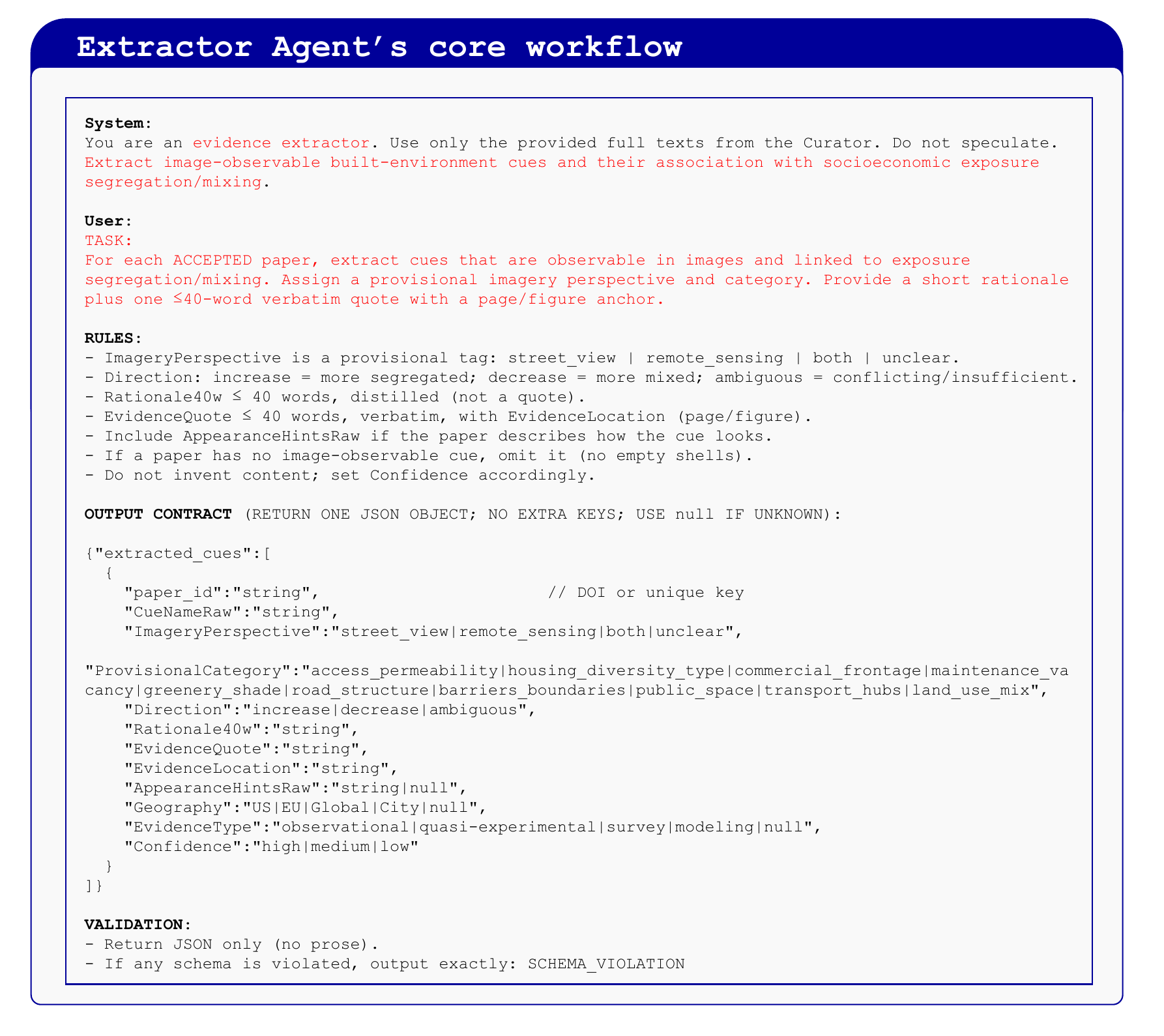}
\caption{Illustration of the agentic workflow of the Extractor agent.} 
\label{fig:S1_2}
\end{figure}

\begin{figure}[htbp]
\centering
\includegraphics[width=0.8\linewidth]{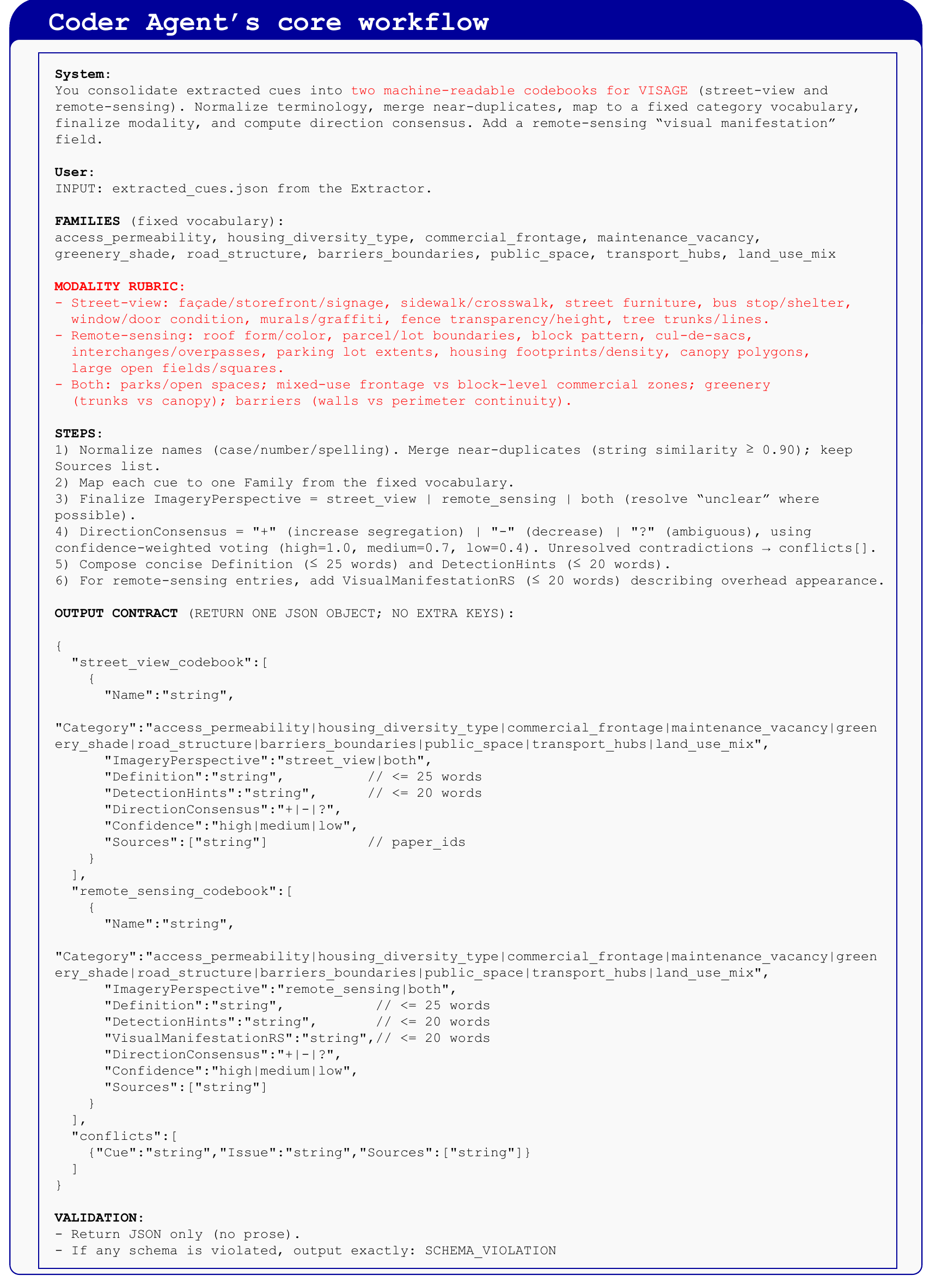}
\caption{Illustration of the agentic workflow of the Coder agent.} 
\label{fig:S1_3}
\end{figure}

\begin{figure}[htbp]
\centering
\includegraphics[width=1.0\linewidth]{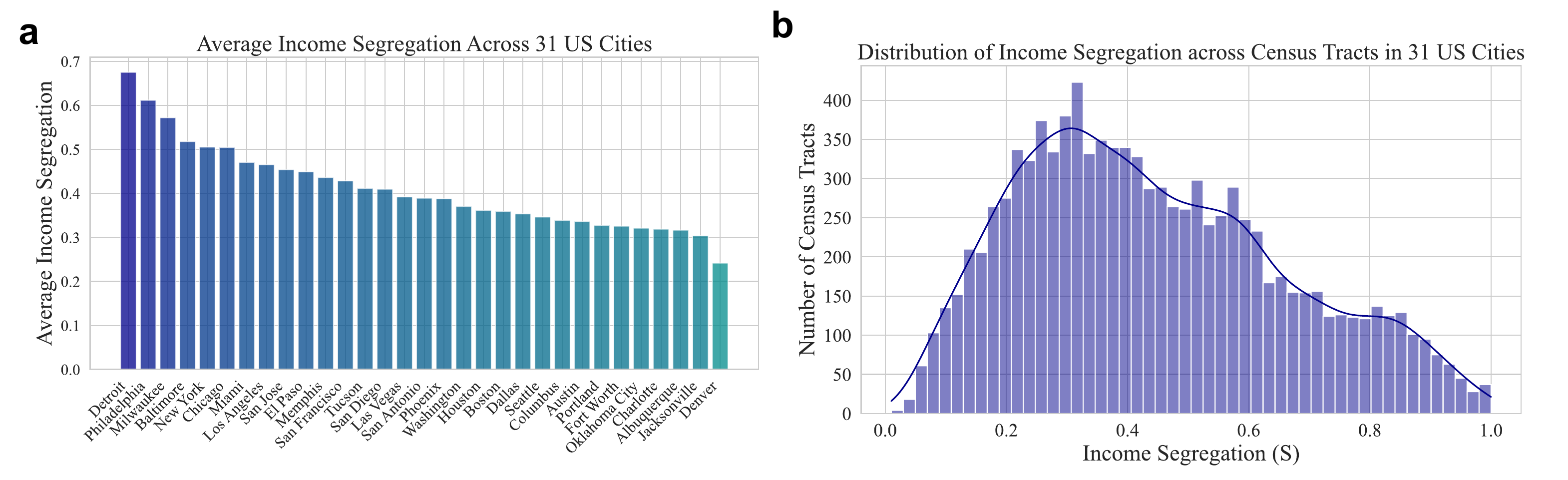}
\caption{Socioeconomic exposure segregation in 31 largest U.S cities. \textbf{a} The average exposure segregation for each city. \textbf{b} The distribution of community-level exposure segregation across the entire 10,030 communities.}
\label{fig:S5}
\end{figure}

\begin{figure*}[ht]
\centering
\includegraphics[width=\linewidth]{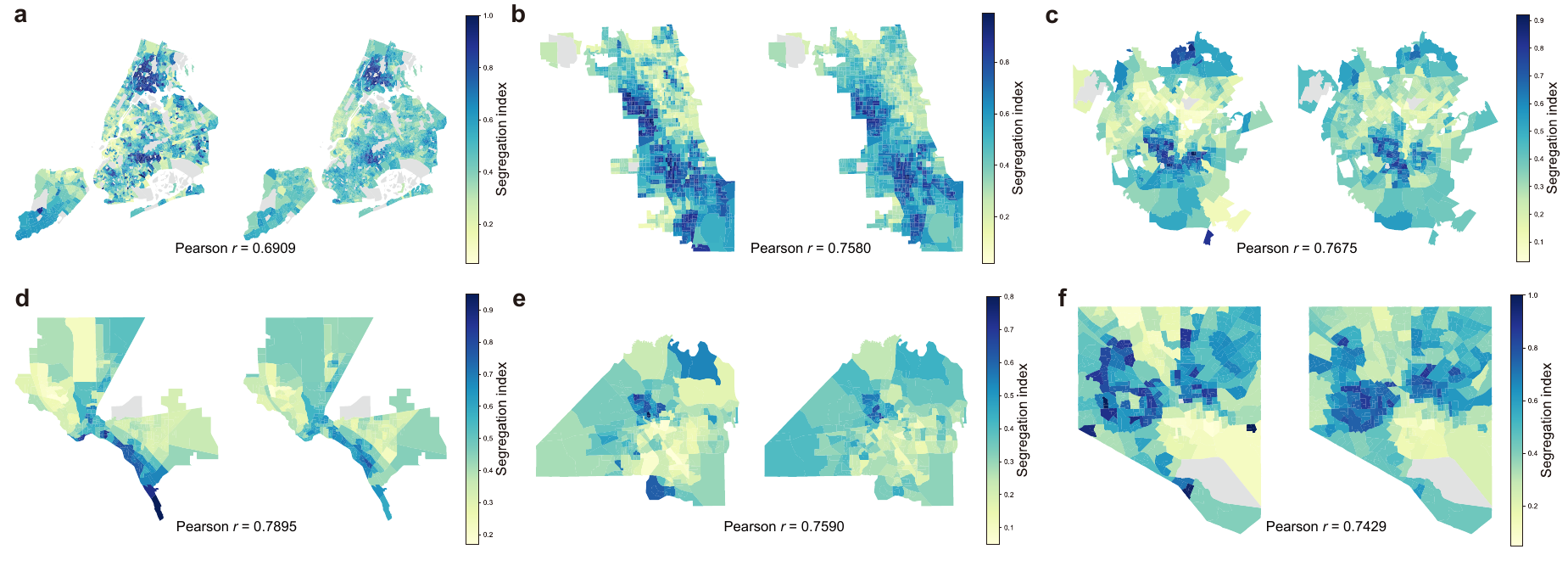}
\caption{Spatial distributions of the ground truth and VISAGE-perceived segregation index for communities in New York (\textbf{a}), Chicago (\textbf{b}), San Antonio (\textbf{c}), El Paso (\textbf{d}), Jacksonville (\textbf{e}), Baltimore (\textbf{f}).}
\label{fig:S8}
\end{figure*}

\begin{figure}[htbp]
\centering
\includegraphics[width=\linewidth]{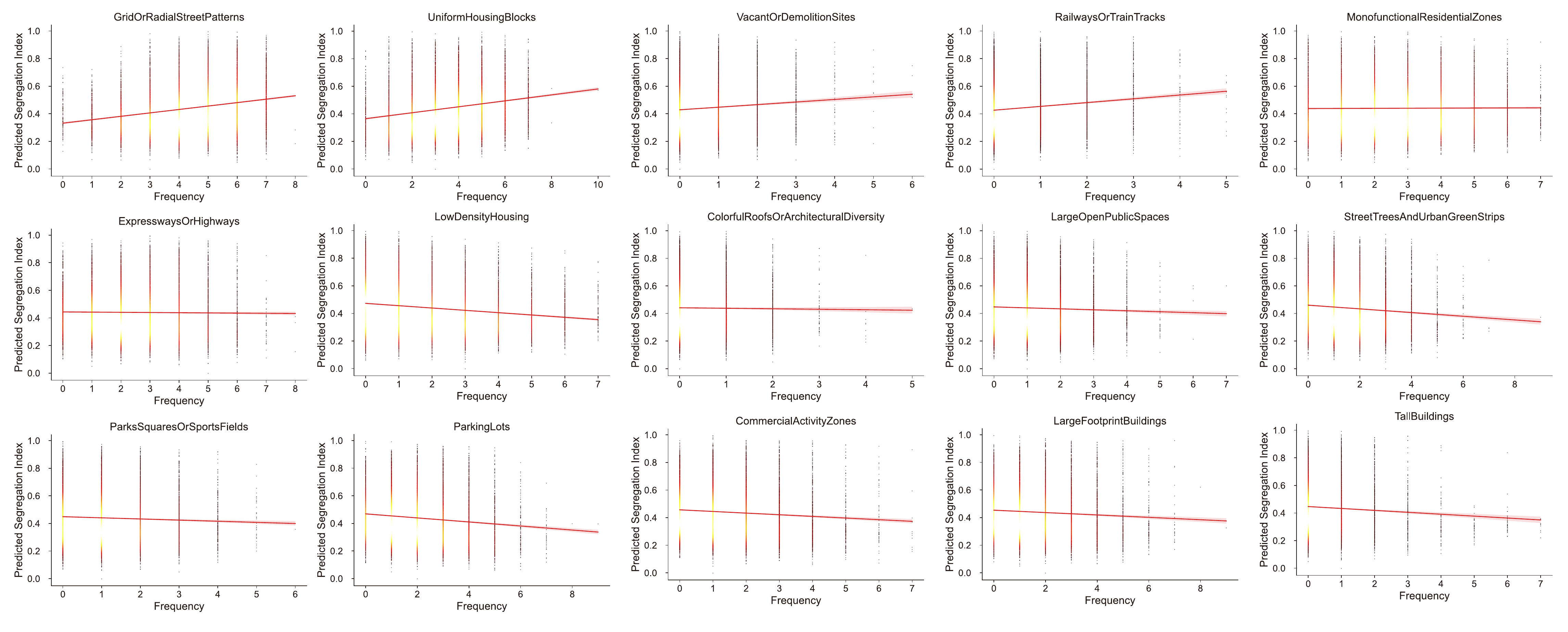}
\caption{Response curves of satellite image visual cues.} 
\label{fig:S9}
\end{figure}

\begin{figure}[htbp]
\centering
\includegraphics[width=1.0\linewidth]{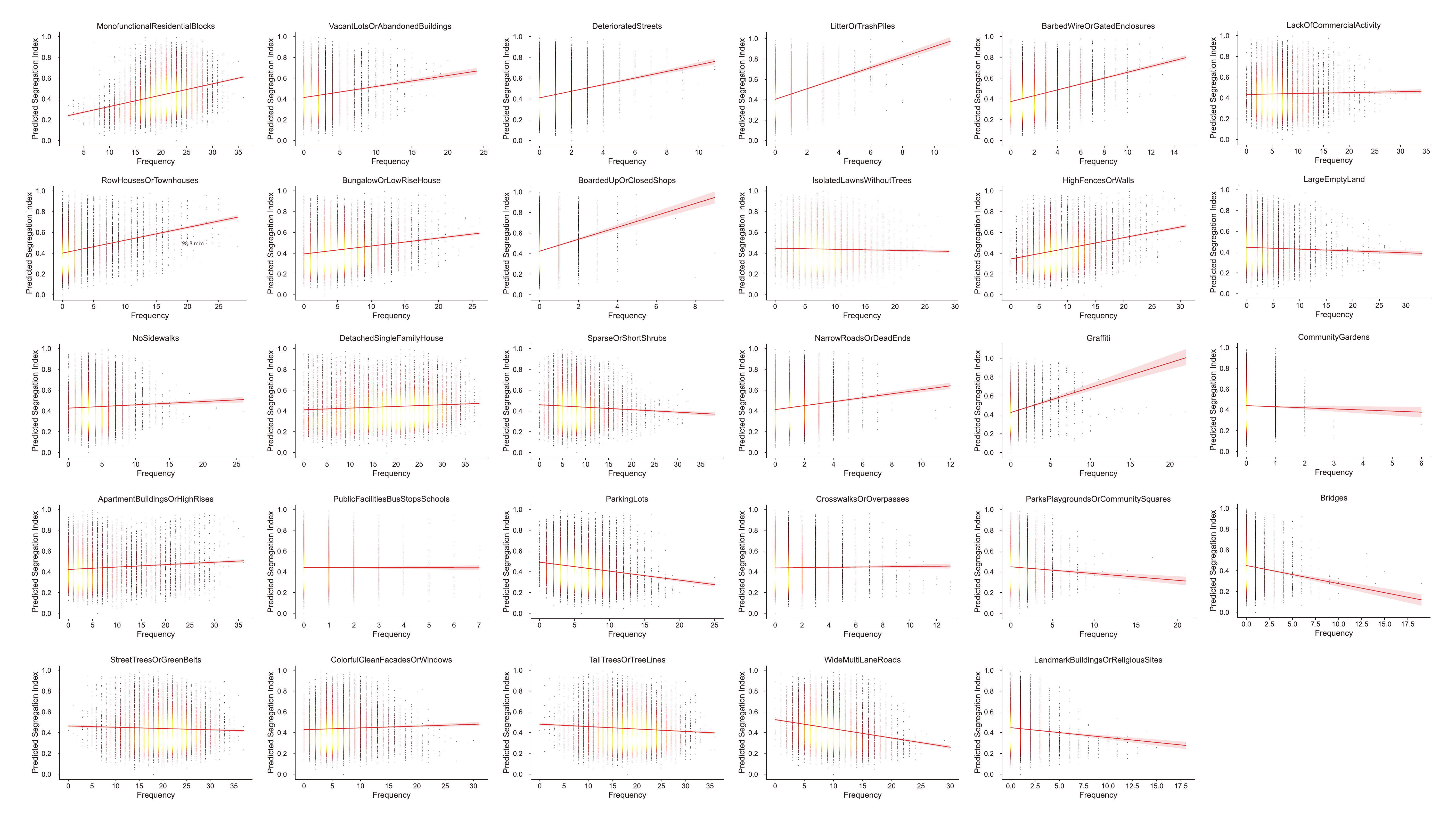}
\caption{Response curves of street-view image visual cues.} 
\label{fig:S10}
\end{figure}

\clearpage
\section*{Supplementary Tables}
\addcontentsline{toc}{section}{Supplementary Tables}

{\small
\begin{longtable}{@{}p{0.03\textwidth} p{0.20\textwidth} p{0.15\textwidth} p{0.10\textwidth} p{0.12\textwidth} p{0.30\textwidth}@{}}
\caption{Retrieved research articles on urban segregation with corresponding visual cues and modalities.} \label{tab:S1} \\

\toprule
\textbf{ID} & \textbf{Title} & \textbf{Publication Info} & \textbf{Modality} & \textbf{Primary Discipline} & \textbf{Visual Cues \& Direction}  \\
\midrule
\endfirsthead

\multicolumn{5}{c}{{\tablename\ \thetable{} -- continued from previous page}} \\
\toprule
\textbf{ID} & \textbf{Title} & \textbf{Publication Info} & \textbf{Modality} & \textbf{Primary Discipline} & \textbf{Visual Cues \& Direction} \\
\midrule
\endhead

\bottomrule
\multicolumn{5}{r}{{Continued on next page}} \\
\endfoot

\bottomrule
\endlastfoot

\cite{Fraser2024TaleOfManyCities} & A tale of many cities: Mapping social infrastructure and social capital across the United States & 2024, Computers, Environment and Urban Systems & Satellite & Urban Planning & CommercialActivityZones(↓), ParksSquaresOrSportsFields(↓), LandmarkBuildingsOrReligiousSites(↓), PublicFacilitiesBusStopsSchools(↓) \\
\hline

\cite{ijerph15050861} & Associations between Urban Sprawl and Life Expectancy in the United States & 2018, International Journal of Environmental Research and Public Health & Satellite & Urban Planning & LowDensityHousing(↑), WideStreetsWithMediansOrShoulders(↑), ExpresswaysOrHighways(↑) \\
\hline

\cite{Yang2023Association} & Association of Neighborhood Racial and Ethnic Composition and Historical Redlining With Built Environment Indicators Derived From Street View Images in the US & 2023, JAMA Network Open & Street-view & Sociology & CrosswalksOrOverpasses(mixed), StreetTreesOrGreenBelts(↓), VacantLotsOrAbandonedBuildings(↑), DetachedSingleFamilyHouse/ApartmentBuildingsOrHighRises(mixed) \\
\hline

\cite{Duncan2014Spatially} & A spatially explicit approach to the study of socio-demographic inequality in the spatial distribution of trees across Boston neighborhoods & 2014, Spatial Demography & Satellite & Environment & StreetTreesOrGreenBelts(mixed), TallTreesOrTreeLines(mixed) \\
\hline

\cite{Roberto2025Constructing} & Constructing Segregation: Examining Social and Spatial Division in Road Networks & 2024, OSF preprint & Satellite & Transportation & CulDeSacsOrFencedBlocks(↑), DisconnectedStreetSegments(↑), ExpresswaysOrHighways/OverpassesOrInterchanges(↑) \\
\hline

\cite{Galeano2023Delivering} & Delivering the 'miracle': levelling-up low-income neighbourhoods through local infrastructure and jobs activation Medellín, Colombia, 2000–2018 & 2023, UCL PhD Thesis & Satellite & Urban Planning & TransportHubsOrStations(↓), ParksSquaresOrSportsFields(↓), OverpassesOrInterchanges(mixed) \\
\hline

\cite{Wurm2018Detecting} & Detecting social groups from space – Assessment of remote sensing-based mapped morphological slums using income data & 2018, Remote Sensing Letters & Satellite & Housing & VacantOrDemolitionSites(↑), UniformHousingBlocks(↑), FragmentedDevelopmentPatches(↑) \\
\hline

\cite{thornton2016disparities} & Disparities in Pedestrian Streetscape Environments by Income and Race Ethnicity & 2016, SSM - Population Health & Street-view & Transportation & Sidewalks/CrosswalksOrOverpasses(↑), ParksPlaygroundsOrCommunitySquares(↑) \\
\hline

\cite{Fan2023Diversity} & Diversity beyond density: Experienced social mixing of urban streets & 2023, PNAS Nexus & Satellite & Sociology & CommercialActivityZones(↓), ConvenienceStoresOrRestaurants(↓), Residential mix proxy(↓) \\
\hline

\cite{Vachuska2024Do} & Do neighborhoods have boundaries? A novel empirical test for a historic question & 2024, PLoS ONE & Satellite & Transportation & ExpresswaysOrHighways(↑), RailwaysOrTrainTracks(↑), Bridges(↑) \\
\hline

\cite{LensMonkkonen2016Do} & Do Strict Land Use Regulations Make Metropolitan Areas More Segregated by Income? & 2016, Journal of the American Planning Association & Satellite & Urban Planning & LowDensityHousing(↑), MonofunctionalResidentialZones(↑) \\
\hline

\cite{Petrovic2019Environmental} & Environmental and social dimensions of community gardens in East Harlem & 2019, Landscape and Urban Planning & Street-view & Environment & CommunityGardens(↓), ParksPlaygroundsOrCommunitySquares(↓) \\
\hline

\cite{BalodeBerzins2025EthnicResidentialRiga} & Ethnic residential patterns in the inner-city core of Riga, Latvia using scalable individualized neighborhoods & 2025, Frontiers in Sustainable Cities & Satellite & Sociology & TallBuildings/UniformHousingBlocks(↑) \\
\hline

\cite{LeGoix2005GatedCommunities} & Gated Communities: Sprawl and Social Segregation in Southern California & 2005, Housing Studies & Satellite & Housing & BarbedWireOrGatedEnclosures/ HighFencesOrWalls(↑), CulDeSacsOrFencedBlocks(↑) \\
\hline

\cite{nilforoshan2023human} & Human mobility networks reveal increased segregation in large U.S. cities & 2023, Nature & Satellite & Sociology & TransportHubsOrStations(mixed), CommercialActivityZones(mixed), LargeFootprintBuildings(mixed) \\
\hline

\cite{Najmi2022Integrating} & Integrating Remote Sensing and Street View Imagery for Mapping Slums & 2022, ISPRS International Journal of Geo-Information & Both (satellite + street-view) & Housing & VacantOrDemolitionSites(↑), UniformHousingBlocks(↑), UnpavedRoadsOrDirtTracks(↑) \\
\hline

\cite{WangVermeulen2021LifeBetweenBuildings} & Life between buildings from a street view image: What do big data analytics reveal about neighbourhood organisational vitality? & 2021, Urban Studies & Street-view & Urban Planning & VisibleShopsOrColorfulSigns(↓), ParksPlaygroundsOrCommunitySquares(↓) \\
\hline

\cite{Moore2024LivingTogetherOrApart} & Living together or apart? Gated condominium communities and social segregation in Bangkok & 2024, Housing Studies & Satellite & Housing & HighFencesOrWalls/ BarbedWireOrGatedEnclosures(↑) \\
\hline

\cite{Vesselinov2008MembersOnly} & Members Only: Gated Communities and Residential Segregation in the U.S. & 2008, Sociological Forum & Satellite & Housing & BarbedWireOrGatedEnclosures/ HighFencesOrWalls(↑) \\
\hline

\cite{Yankey2024NeighborhoodRacialSegregation} & Neighborhood Racial Segregation Predicts the Spatial Distribution of Supermarkets and Grocery Stores & 2023, Journal of Racial and Ethnic Health Disparities & Satellite &  Sociology & CommercialActivityZones(↑), LackOfCommercialActivity(↑) \\
\hline

\cite{BabereChingwele2018NeighbourhoodUnitResidentialSegregation} & Neighbourhood Unit Residential Segregation in the Global South and Its Impact on Settlement Development: The Case of Dar es Salaam City & 2018, International Journal of Scientific Research & Satellite & Urban Planning & UniformHousingBlocks(↑), UnpavedRoadsOrDirtTracks(↑) \\
\hline

\cite{Serrano2023PedestrianOriented} & Pedestrian-oriented zoning moderates the relationship between racialized economic segregation and active travel to work, United States & 2023, Preventive Medicine & Satellite (zoning/transport) & Transportation & CrosswalksOrOverpasses/ PublicFacilitiesBusStopsSchools(↓), GridOrRadialStreetPatterns(↓) \\
\hline

\cite{Marineau2025RacializedEconomicSegregation} & Racialized economic segregation, the built environment, and assault-related injury: Moderating role of green space and vacant housing & 2025, Preventive Medicine Reports & Satellite (land cover/parks, vacancy) & Sociology & ParksSquaresOrSportsFields/ StreetTreesOrGreenBelts(↓), VacantOrDemolitionSites(↑) \\
\hline

\cite{Locke2021ResidentialHousingSegregation} & Residential housing segregation and urban tree canopy in 37 US Cities & 2021, npj Urban Sustainability & Satellite & Environment & StreetTreesOrGreenBelts/ TallTreesOrTreeLines(↑) \\
\hline

\cite{KwateLohWhiteSaldana2013RetailRedlining} & Retail Redlining in New York City: Racialized Access to Day-to-Day Retail Services & 2013, Journal of Urban Health & Satellite (POI) & Sociology & LackOfCommercialActivity(↑), BoardedUpOrClosedShops(↑), VisibleShopsOrColorfulSigns(↓) \\
\hline

\cite{KnaapRey2024SegregatedByDesign} & Segregated by design? Street network topological structure and the measurement of urban segregation & 2024, Environment and Planning B & Satellite (network topology/OSM) & Urban Planning & LowStreetConnectivity/ DisconnectedStreetSegments(↑), CulDeSacsOrFencedBlocks(↑), ExpresswaysOrHighways(↑) \\
\hline

\cite{RogersGardnerCarlson2013SocialCapitalWalkability} & Social Capital and Walkability as Social Aspects of Sustainability & 2013, Sustainability & Street-view & Sociology & Sidewalks/CrosswalksOrOverpasses(↓), VisibleShopsOrColorfulSigns(↓), ParksPlaygroundsOrCommunitySquares(↓) \\
\hline

\cite{AbitbolKarsai2020SocioeconomicCorrelations} & Socioeconomic correlations of urban patterns inferred from aerial images: interpreting activation maps of Convolutional Neural Networks & 2020, preprint & Satellite & Urban Planning & UniformHousingBlocks(↑), LargeFootprintBuildings(mixed), GridOrRadialStreetPatterns(mixed) \\
\hline

\cite{Useche2024SpatialSegregationColombianCities} & Spatial Segregation Patterns and Association With Built Environment Features in Colombian Cities & 2024, Cities & Satellite & Housing & CulDeSacsOrFencedBlocks(↑), MonofunctionalResidentialZones(↑), LowStreetConnectivity(↑) \\
\hline

\cite{Odgers2012SystematicSocialObservation} & Systematic social observation of children's neighborhoods using Google Street View: a reliable and cost-effective method & 2012, Journal of Child Psychology and Psychiatry & Street-view (method validation) & Sociology & Graffiti(↑), LitterOrTrashPiles(↑), DeterioratedStreets(↑) \\
\hline

\cite{Ananat2011WrongSides} & The wrong side(s) of the tracks: The causal effects of racial segregation on urban poverty and inequality & 2011, American Economic Journal: Applied Economics & Satellite & Sociology & RailwaysOrTrainTracks(↑) \\
\hline

\cite{RothwellMassey2009DensityZoningSegregation} & The Effect of Density Zoning on Racial Segregation in U.S. Urban Areas & 2009, Urban Affairs Review & Satellite (zoning/land use) & Housing & LowDensityHousing(↑), MonofunctionalResidentialZones(↑) \\
\hline

\cite{FraserYabeAldrichMoro2024GreatEqualizer} & The great equalizer? Mixed effects of social infrastructure on diverse encounters in cities & 2024, Computers, Environment and Urban Systems & Satellite (mobility + POI) & Sociology & ParksSquaresOrSportsFields(↓), CommercialActivityZones(↓), CommunityFacilities(mixed) \\
\hline

\cite{SalazarMiranda2020ShapeOfSegregation} & The Shape of Segregation: The Role of Urban Form in Immigrant Assimilation & 2020, Cities & Satellite (urban form metrics) & Urban Planning & GridOrRadialStreetPatterns(↓), LowStreetConnectivity(↑) \\
\hline

\cite{Roberto2018SpatialProximityConnectivity} & The Spatial Proximity and Connectivity Method for Measuring and Analyzing Residential Segregation & 2018, Sociological Methodology & Satellite (road-network-aware metrics) & Housing & NarrowRoadsOrDeadEnds/ CulDeSacsOrFencedBlocks/ ExpresswaysOrHighways(↑) \\
\hline

\cite{RobertoKorverGlenn2021SpatialStructureLocalExperience} & The Spatial Structure and Local Experience of Residential Segregation & 2021, Spatial Demography & Satellite (barriers/connectivity) & Sociology & DisconnectedStreetSegments(↑), ExpresswaysOrHighways(↑), RailwaysOrTrainTracks(↑) \\
\hline

\cite{LiangWang2019UncoveringDominantSocialClass} & Uncovering Dominant Social Class in Neighborhoods through Building Footprints: A Case Study of Residential Zones in Massachusetts using Computer Vision & 2019, preprint & Satellite (building footprints) & Housing & UniformHousingBlocks(↑), LargeFootprintBuildings(mixed) \\
\hline

\cite{Aiello2025UrbanHighwaysBarriers} & Urban highways are barriers to social ties & 2025, PNAS & Satellite (infrastructure + social graph) & Transportation & ExpresswaysOrHighways(↑), OverpassesOrInterchanges(↑), WideMultiLaneRoads(↑) \\
\hline

\cite{SampsonSchachnerMare2017UrbanIncomeInequality} & Urban Income Inequality and the Great Recession in Sunbelt Form: Disentangling Individual and Neighborhood-Level Change in Los Angeles & 2017, RSF: The Russell Sage Foundation Journal of the Social Sciences & Census/context (not imagery) & Sociology & LowDensityHousing/Urban Form proxies(↑) \\
\hline

\cite{Figueroa2019UrbanStructureSocialSegregation} & Urban structure and the layout of social segregation & 2019, 12th International Space Syntax Symposium & Satellite & Urban Planning & GridOrRadialStreetPatterns(↓), LowStreetConnectivity/DisconnectedStreetSegments(↑) \\
\hline

\cite{Zhang2024UrbanVisualIntelligence} & Urban Visual Intelligence: Studying Cities with Artificial Intelligence and Street-Level Imagery & 2024, Annals of the American Association of Geographers & Street-view (review) & Urban Planning & Multiple; summary article (no single direction) \\
\hline

\cite{GebruKrauseWangChenDengAidenFeiFei2017DeepLearningStreetView} & Using Deep Learning and Google Street View to Estimate the Demographic Makeup of Neighborhoods Across the United States & 2017, PNAS & Street-view & Sociology & VisibleStreetscape/Car types(↑) \\
\hline

\cite{HemerijckxVanEmelenRymenantsDavisVerburgLwasaVanRompaey2020Upscaling} & Upscaling Household Survey Data Using Remote Sensing to Map Socioeconomic Groups in Kampala, Uganda & 2020, Remote Sensing & Satellite & Housing & UniformHousingBlocks(↑), UnpavedRoadsOrDirtTracks(↑), FragmentedDevelopmentPatches(↑) \\
\hline

\cite{Marco2017ValidationGSVNeighborhoodDisorder} & Validation of a Google Street View-Based Neighborhood Disorder Observational Scale & 2017, Journal of Urban Health & Street-view & Sociology & Graffiti(↑), LitterOrTrashPiles(↑), DeterioratedStreets(↑), BoardedUpOrClosedShops(↑) \\

\end{longtable}
}

\begin{table}[ht]
\centering
\small
\caption{Satellite imagery visual cue codebook}
\begin{tabular}{p{0.30\linewidth}p{0.30\linewidth}p{0.30\linewidth}}
\toprule
\textbf{Visual cue} & \textbf{Visual cue} & \textbf{Visual cue} \\
\midrule
Colorful Roofs or Architectural Diversity & Commercial Activity Zones & Cul-de-sacs or Fenced Blocks \\
Disconnected Street Segments & Expressways or Highways & Fragmented Development Patches \\
Grid or Radial Street Patterns & Isolated Large Lots & Large Footprint Buildings \\
Large Open Public Spaces & Low Density Housing & Low Street Connectivity \\
Monofunctional Residential Zones & Overpasses or Interchanges & Parking Lots \\
Parks Squares or Sports Fields & Railways or Train Tracks & Scattered Lawns or Private Greenery \\
Schools or Educational Campuses & Street Trees and Urban Green Strips & Tall Buildings \\
Transport Hubs or Stations & Uniform Housing Blocks & Unpaved Roads or Dirt Tracks \\
Vacant or Demolition Sites & Wide Streets with Medians or Shoulders & \\
\bottomrule
\end{tabular}
\label{tab:S2}
\end{table}

\begin{table}[ht]
\centering
\small
\setlength{\tabcolsep}{6pt}
\caption{Street-view visual cue codebook.}
\label{fig:sv_codebook}
\begin{tabular}{p{0.23\linewidth}p{0.23\linewidth}p{0.23\linewidth}p{0.23\linewidth}}
\toprule
\textbf{Visual cue} & \textbf{Visual cue} & \textbf{Visual cue} & \textbf{Visual cue} \\
\midrule
Apartment Buildings or High-Rises & Barbed Wire or Gated Enclosures & Boarded-Up or Closed Shops & Bridges \\
Bungalow or Low-Rise House & Colorful Clean Facades or Windows & Community Gardens & Convenience Stores or Restaurants \\
Crosswalks or Overpasses & Detached Single-Family House & Deteriorated Streets & Graffiti \\
High Fences or Walls & Isolated Lawns Without Trees & Lack of Commercial Activity & Landmark Buildings or Religious Sites \\
Large Empty Land & Litter or Trash Piles & Monofunctional Residential Blocks & Murals, Sculptures, or Urban Art \\
Narrow Roads or Dead Ends & No Sidewalks & Parking Lots & Parks, Playgrounds, or Community Squares \\
Public Facilities, Bus Stops, or Schools & Row Houses or Townhouses & Sparse or Short Shrubs & Street Trees or Green Belts \\
Tall Trees or Tree Lines & Vacant Lots or Abandoned Buildings & Visible Shops or Colorful Signs & Wide Multi-Lane Roads \\
\bottomrule
\end{tabular}
\label{tab:S3}
\end{table}

\begin{table*}[t]
\centering
\small
\caption{OLS estimates for ground-truth and predicted exposure segregation (coefficients with 95\% CIs). Significance: $^{***}p<0.001$, $^{**}p<0.01$, $^{*}p<0.05$.}
\label{tab:S4}
\setlength{\tabcolsep}{8pt}
\begin{tabular}{lcccc}
\toprule
& \multicolumn{2}{c}{\textbf{(1) Ground-truth $S$}} & \multicolumn{2}{c}{\textbf{(2) Predicted $\hat S$}} \\
\cmidrule(lr){2-3}\cmidrule(lr){4-5}
\textbf{Variable} & \textbf{Coef.} & \textbf{95\% CI} & \textbf{Coef.} & \textbf{95\% CI} \\
\midrule
\multicolumn{5}{l}{\textit{Inclusionary Housing}} \\
\textbf{Is Inclusionary Housing} & \textbf{$-0.0684^{***}$} & \textbf{[$-0.0978$, $-0.0390$]} & \textbf{$-0.0482^{***}$} & \textbf{[$-0.0702$, $-0.0262$]} \\
\addlinespace[0.3em]
\multicolumn{5}{l}{\textit{Demographics}} \\
Asian Ratio & $0.4974^{***}$ & [$0.3622$, $0.6326$] & $0.4467^{***}$ & [$0.3467$, $0.5467$] \\
Hispanic Ratio & $0.4654^{***}$ & [$0.3400$, $0.5908$] & $0.3946^{***}$ & [$0.3005$, $0.4887$] \\
Non-Hispanic White Ratio & $0.3662^{***}$ & [$0.2368$, $0.4956$] & $0.3041^{***}$ & [$0.2061$, $0.4021$] \\
Non-Hispanic Black Ratio & $0.5946^{***}$ & [$0.4672$, $0.7220$] & $0.5182^{***}$ & [$0.4222$, $0.6142$] \\
\addlinespace[0.3em]
\multicolumn{5}{l}{\textit{Socioeconomic Controls}} \\
Median Income & $2.518\times10^{-6\,***}$ & [$2.298\times10^{-6}$, $2.738\times10^{-6}$] & $1.765\times10^{-6\,***}$ & [$1.602\times10^{-6}$, $1.928\times10^{-6}$] \\
Bachelor Ratio & $-0.2844^{***}$ & [$-0.3158$, $-0.2530$] & $-0.2413^{***}$ & [$-0.2648$, $-0.2178$] \\
Poverty Ratio & $0.4391^{***}$ & [$0.3979$, $0.4803$] & $0.3415^{***}$ & [$0.3101$, $0.3729$] \\
\addlinespace[0.3em]
\multicolumn{5}{l}{\textit{Constant}} \\
Constant & $-0.1493^{*}$ & [$-0.2747$, $-0.0239$] & $-0.0367$ & [$-0.1288$, $0.0554$] \\
\midrule
Observations & \multicolumn{2}{c}{10,030} & \multicolumn{2}{c}{10,030} \\
$R^{2}$ & \multicolumn{2}{c}{0.238} & \multicolumn{2}{c}{0.317} \\
Adj. $R^{2}$ & \multicolumn{2}{c}{0.237} & \multicolumn{2}{c}{0.317} \\
F-statistic & \multicolumn{2}{c}{390.4} & \multicolumn{2}{c}{582.4} \\
\bottomrule
\end{tabular}
\end{table*}

%TC:ignore

\section*{Data availability}
Street-level imagery was retrieved via the Google Maps Street View API (https://developers.google.com/maps/documentation/streetview/overview). Satellite imagery is Esri USA Satellite 2019 (https://learn.arcgis.com/en/projects/download-imagery-from-an-online-database/). Community boundaries follow U.S. Census Bureau TIGER/Line tracts. Inclusionary housing data are from the Inclusionary Housing Map \& Program Database (https://inclusionaryhousing.org/map/). The SafeGraph Weekly Patterns datasets used to construct ground-truth exposure indices are available from SafeGraph. Access is restricted under SafeGraph’s Data License Agreement, and these data are licensed to the authors and not publicly available. Demographic data are publicly available from the U.S. Decennial Census and American Community Survey at (https://www.safegraph.com/free-data/open-census-data).

\section*{Code availability}
All code used for agent workflow, data process, model tuning, and figure generation in this study is available via GitHub at https://github.com/tsinghua-fib-lab/VISAGE.

\section*{Author contributions}
Y.L. launched this project, provided the research outline and guided the overall research direction. Y.Z., X.Z., F.X., T.X., and Y. Li contributed ideas. Y.Z. and R.M. designed the method, conducted experiments, and prepared figures, with support from X.Z. and Y.S. on large multi-modal model training and fine-tuning. All authors jointly participated in the writing of the paper.

\section*{Competing interests}
The authors declare no competing interests.

%TC:endignore

\bibliography{reference}

\end{document}